\documentclass[review]{elsarticle}

\usepackage[utf8]{inputenc}	
\usepackage[T1]{fontenc}	
\usepackage{CJKutf8}	

\usepackage{lineno,hyperref}
\modulolinenumbers[5]

\usepackage{booktabs}
\usepackage{makecell}	
\usepackage{multirow}   
\usepackage{natbib}	
\usepackage{orcidlink}	
\usepackage{subcaption}	
\usepackage{tabularx}	
\usepackage{utfsym}	
\usepackage{adjustbox}
\usepackage{array}

\usepackage{geometry}
\usepackage{setspace}
\usepackage[flushleft]{threeparttable}

\usepackage{fancyhdr}
\usepackage[section]{placeins}	

\usepackage{listings}

\lstdefinestyle{promptstyle}{
    basicstyle=\ttfamily\footnotesize,
    breaklines=true,
    breakatwhitespace=true,
    breakindent=0pt,
    breakautoindent=false,
    columns=fullflexible,
    keepspaces=true,
    showstringspaces=false,
    frame=single,
    numbers=none,
    xleftmargin=0.5em,
    xrightmargin=0.5em,
    aboveskip=0.8em,
    belowskip=0.8em
}

\biboptions{authoryear}

\renewcommand{\thefootnote}{\fnsymbol{footnote}}
\makeatletter
\def\thefootnote{\fnsymbol{footnote}}
\makeatother

\begin{document}
\begin{CJK}{UTF8}{gbsn}	

\renewcommand{\thefootnote}{\fnsymbol{footnote}} 
\setcounter{footnote}{1} 

\begin{frontmatter}


\title{FMMD: A multimodal multidisciplinary dataset of open peer reviews from F1000Research}



\author{Zhenzhen Zhuang$\rm^{a,c}$, Yuqing Fu$\rm^{a,c,\dagger}$, Jing Zhu$\rm^{a,c,\dagger}$, Zhangping Zhou$\rm^{b}$, Jialiang Lin$\rm^{a,c,*}$}

\cortext[mycorrespondingauthor]{Corresponding author\\
\hspace*{0.42cm}$^{\dagger}$Equal contribution\\
\hspace*{0.5cm}\textit{Email addresses:} zhuangzz@gzist.edu.cn (Z. Zhuang), zhouzp@xmu.edu.cn (Z. Zhou), me@linjialiang.net (J. Lin)
}

\address[Aaddress]{School of Computer Science and Engineering, Guangzhou Institute of Science and Technology, Guangzhou, China}
\address[Baddress]{College of Foreign Languages and Cultures, Xiamen University, Xiamen, China}
\address[Caddress]{Science and Education Evaluation Lab, Guangzhou Institute of Science and Technology, Guangzhou, China}

\begin{abstract}

Automated scholarly paper review (ASPR) has entered the coexistence phase with traditional peer review, where artificial intelligence (AI) systems are increasingly incorporated into real-world manuscript evaluation. In parallel, research on automated and AI-assisted peer review has proliferated. Despite this momentum, empirical progress remains constrained by several critical limitations in existing datasets. While reviewers routinely evaluate figures, tables, and complex layouts to assess scientific claims, most existing datasets remain overwhelmingly text-centric. This bias is reinforced by a narrow focus on data from computer science publications. Furthermore, existing datasets rarely preserve precise alignment between review comments and specific manuscript versions, obscuring the iterative relationship between peer review and manuscript evolution. In response, we introduce FMMD, a multimodal and multidisciplinary open peer review dataset curated from F1000Research. The dataset addresses the current limitations by integrating manuscript-level visual and structural data with version-specific reviewer reports and editorial decisions. By explicitly aligning review comments with the exact article version under review, FMMD enables granular analysis of the peer review lifecycle. Importantly, its coverage of F1000Research extends ASPR research beyond its traditional focus on computer science to a diverse range of scientific disciplines. FMMD supports a range of research tasks, including visual-semantic consistency classification, figure-related review comment generation, and editorial decision prediction based on multimodal manuscript inputs, thereby providing a comprehensive empirical resource for developing and evaluating multimodal ASPR systems and advancing peer review research.
\end{abstract}


\end{frontmatter}


\section{Introduction}
\label{sec:intro}

Scholarly papers are inherently complex information carriers composed of multiple interacting modalities, including text, tables, figures, and layout structure~\citep{cai-uni-2024}. During the writing process, authors convey research motivations, problem formulations, and methodological contributions through multimodal collaboration. Textual narratives establish formal definitions and conceptual grounding, while figures and tables visually summarize workflows, experimental settings, and key findings. Building on these elements, structured layouts guide readers through the logical organization of the work and facilitate efficient access to core content~\citep{jewitt-introducing-2025}. These modalities operate in a complementary manner, collectively enabling the coherent and effective communication of scientific contributions. Reliance on a single modality often fails to adequately capture the full scope, structure, and internal logic of a study.

Multimodal information is equally central to the academic peer review process. When evaluating a manuscript, reviewers typically engage first with visual components such as tables and figures before conducting a detailed reading of the main text. Clear graphical representations, well-organized paragraphs, and coherent layout design substantially reduce cognitive load and improve information retrieval efficiency, enabling reviewers to more effectively assess a paper's technical quality, novelty, and clarity~\citep{jung-what-2025,acm-evaluating-2025}. Peer review is therefore not a purely textual activity but rather an integrated cognitive process grounded in multimodal perception and reasoning~\citep{gao-mmreview-2025}.

Automated scholarly paper review (ASPR)~\citep{lin-automated-2023} should consequently aspire to evaluate scientific manuscripts with a level of multimodal comprehensiveness comparable to that of human reviewers. This concept, which we previously proposed, has gained traction within the research community as a solution to the long-standing limitations of traditional peer review, including low efficiency, outcome variability, and restricted domain coverage~\citep{staudinger-analysis-2024,couto-relevai-2024,du-llms-2024}.

Despite ongoing caution from academia and publishers regarding artificial intelligence (AI)-assisted peer review~\citep{elsevier-use-2024,ieee-become-2024,springer-editorial-2024,cambridge-core-2024}, generative AI has already been widely integrated into scholarly workflows~\citep{eger-transforming-2025}. A recent survey reported that about 33\% of researchers have employed AI to assist in peer review tasks~\citep{ng-attitudes-2025}. In parallel, major academic conferences have initiated controlled explorations of AI-supported review. For instance, ICLR 2025 launched the Review Feedback Agent experiment, which leverages multiple large language models (LLMs) to generate feedback for reviewers with the goal of improving review quality and actionability~\citep{thakkar-large-2026}. Similarly, AAAI-26 introduced the AI-Assisted Peer Review Pilot Program, leveraging LLMs to generate draft comments that support human reviewers~\citep{aaai-instructions-2025}. AI systems are now actively integrated into manuscript evaluation, marking a phase where ASPR coexists with traditional peer review~\citep{lin-automated-2023,zhuang-large-2025}.

In response, recent research efforts have sought to enhance the performance of ASPR. Some studies focus on improving LLMs' ability to comprehend scholarly papers~\citep{li-scilitllm-2024,couto-relevai-2024,faizullah-limgen-2024}, while others aim to improve the quality of LLM-generated review comments~\citep{gao-reviewer2-2024,yu-automated-2024,ni-chatreviewer-2023,grag-revieweval-2025}. Although these approaches have achieved notable progress, they remain largely text-centric. As highlighted in our prior work~\citep{zhuang-large-2025}, the emergence of multimodal LLMs has laid the technical groundwork for processing multimodal scientific documents, but systematic research on leveraging multimodal inputs within ASPR remains limited. Current efforts incorporating non-textual elements, such as formulas, tables, or figures, are still largely exploratory and narrow in scope~\citep{taechoyotin-mamorx-2024,markhasin-ai-2025,markhasin-llm-2025}. This overreliance on textual information creates a fundamental mismatch between ASPR and traditional peer review. Consequently, text-centric ASPR falls short of faithfully modeling real-world peer review practices, thereby constraining their practical applicability and undermining their credibility in academic evaluation.

From a technical perspective, this limitation does not primarily stem from deficiencies in model architectures or reasoning capabilities. Recent advances in multimodal LLMs have demonstrated the methodological feasibility of jointly modeling multiple information modalities, including text and vision~\citep{yin-survey-2024,cai-uni-2024,song-how-2025}. On the contrary, the primary impediment to multimodal ASPR stems directly from the data layer. The persistent absence of high-quality, well-structured, and richly annotated multimodal review datasets substantially limits models' ability to systematically capture cross-modal correspondences and to reason about the relationship between multimodal manuscript content and review comments~\citep{zhuang-large-2025}. This challenge is not unique to ASPR but reflects a broader issue in deep learning, where the upper bound of model performance is fundamentally constrained by the scale, structure, and quality of available data. In multimodal settings, data construction is further complicated by the need for accurate alignment and contextual association across modalities, making dataset quality the critical bottleneck constraining model performance.

At the representation level, existing ASPR datasets predominantly rely on plain text, LaTeX, or PDF, yet none provides a complete representation of the manuscript as both a structured and multimodal artifact. Plain text is readily processed but discards multimodal information and explicit document structure. While LaTeX preserves rich structural and mathematical information, its missing source files, macro variability, and complex compilation dependencies hinder reliable processing~\citep{tan-inconsistencies-2024}. PDF, by contrast, retains the visual fidelity of the final manuscript but is inherently difficult to parse computationally, making the reliable extraction of structured elements such as equations, tables, and references a major bottleneck~\citep{ji-aceparse-2025}.

To address these challenges, we construct an HTML-based multimodal review dataset designed to systematically preserve multimodal elements and underlying document structure while optimizing machine readability and scalability. Our primary contributions are as follows:

\begin{itemize}

\item We design and publicly release FMMD, a large-scale, multimodal, and multidisciplinary open peer review dataset based on the F1000Research platform that preserves full-text articles with embedded visual content, peer review comments, and author responses.

\item We establish an explicit cross-modal alignment mechanism within FMMD, directly linking peer review comments and author responses to the corresponding multimodal content in the source articles.

\item We evaluate FMMD on downstream tasks to assess the capabilities of current multimodal LLMs in ASPR and establishing baselines to guide future research.

\end{itemize}

\begin{figure}[htb]
    \centering
    \includegraphics[width=0.96\textwidth]{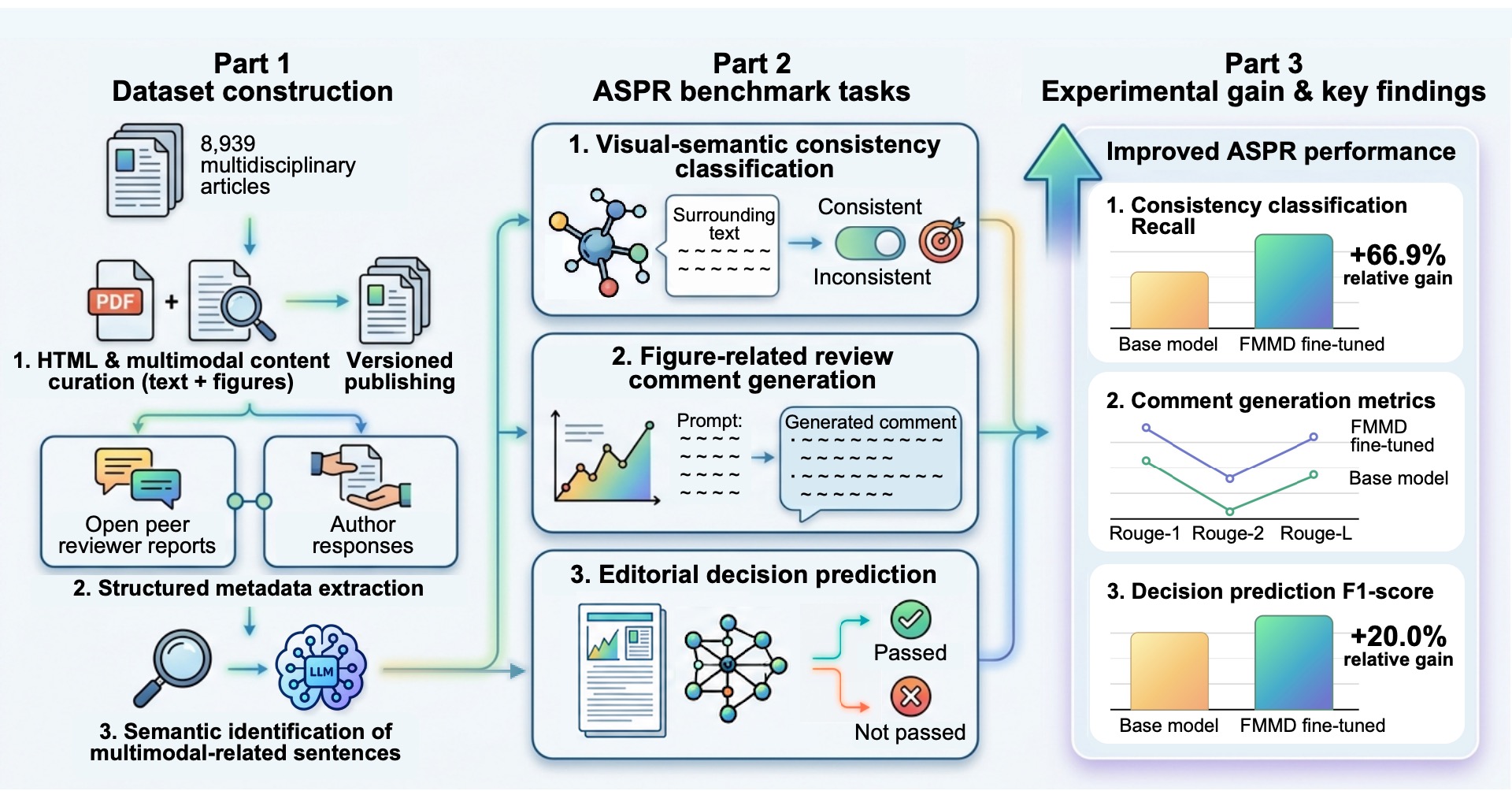}
    \caption{Schematic overview of FMMD}
    \label{fig:schematic-overview}
\end{figure}

The remainder of this paper is organized as follows. Section~\ref{sec:related-work} reviews the related literature on existing multimodal datasets for ASPR. Section~\ref{sec:dataset} details the data construction pipeline, multimodal content curation and alignment mechanism, and statistical properties of the proposed FMMD dataset. Section~\ref{sec:experiments} presents three downstream ASPR benchmark tasks built upon FMMD, including visual-semantic consistency classification, figure-related review comment generation, and editorial decision prediction, along with their corresponding empirical results and performance comparisons against the base model. Finally, Section~\ref{sec:concl} concludes the paper. An end-to-end overview of our dataset construction workflow, benchmark formulations, and key experimental findings is illustrated in Fig.~\ref{fig:schematic-overview}.

\section{Related work}
\label{sec:related-work}

Numerous public datasets of scholarly papers have been released to support diverse research objectives. From the perspective of content composition, existing datasets can be categorized into three types: datasets that include textual content with peer review comments, datasets that include full multimodal content without peer review comments, and datasets that include both multimodal content and peer review comments.

As reviewed in our prior survey~\citep{zhuang-large-2025}, the majority of publicly available datasets for ASPR fall into the first type, providing only textual paper content paired with peer review comments. These datasets have been adopted across different stages of the ASPR pipeline and typically include paper metadata, full-text content, and corresponding review comments. Table~\ref{tab:comment-datasets} summarizes representative scholarly paper datasets from 2025 onward, complementing those covered in our prior survey. Recent efforts have increasingly moved beyond conventional paper-review pairs toward richer and more structured review resources, incorporating LLM-generated reviews, aspect-level annotations, identified research limitations, and review-quality information. This evolution has broadened their applications from conventional review comment generation and editorial decision prediction to human-LLM review comparison, review-quality assessment, fine-grained review analysis, and targeted feedback generation. Despite these advances in scale and structural richness, existing datasets remain predominantly text-centric and largely discard the multimodal content of the source papers. More importantly, they generally provide no explicit annotations linking review comments to multimodal content. Addressing this gap requires datasets that retain original visual elements while establishing explicit links between review comments and specific paper components to which they refer.

\begin{table}[h]
    \centering
    \caption{Representative scholarly paper datasets containing textual content and peer review comments}
    \label{tab:comment-datasets}

    \setcellgapes{1.5pt}
    \makegapedcells
    \small
    
    \begin{threeparttable}

    \begin{adjustbox}{width=1.0\textwidth,keepaspectratio}

    \renewcommand{\arraystretch}{1.5}

    \begin{tabular}{
        >{\raggedright\arraybackslash}p{0.12\textwidth}  
        >{\raggedright\arraybackslash}p{0.11\textwidth}  
        >{\raggedright\arraybackslash}p{0.06\textwidth}  
        >{\raggedright\arraybackslash}p{0.28\textwidth}  
        >{\raggedright\arraybackslash}p{0.17\textwidth}  
        >{\raggedright\arraybackslash}p{0.27\textwidth}  
    }
    \toprule
    
    Name & Source & Domain & Content & Scale & Application \\ \midrule

    \makecell[l]{Dataset of \\Li et al.\\ \citeyear{li-unveiling-2025}} &
    \makecell[l]{OpenReview} &
    \makecell[l]{AI, ML} &
    \makecell[l]{Full papers, \\human reviews, \\LLM reviews} &
    \makecell[l]{1,683 papers,\\6,495 human reviews, \\8,415 LLM reviews} &
    \makecell[l]{Human--LLM review comparison,\\ review quality assessment,\\ paper quality sensitivity evaluation} \\

    \makecell[l]{Dataset of \\Lu et al.\\ \citeyear{lu-identifying-2025}} &
    \makecell[l]{OpenReview} &
    \makecell[l]{NLP, ML} &
    \makecell[l]{Full papers, human reviews, \\ aspect-annotated review sentences \\} &
    \makecell[l]{350 papers,\\ 1,094 reviews,\\ 25k+ aspect \\annotations} &
    \makecell[l]{Review quality assessment,\\ human--LLM review comparison, \\AI-generated review detection} \\

    \makecell[l]{DeepReview-13K\\ \citeyear{zhu-deepreview-2025}} &
    \makecell[l]{OpenReview, \\arXiv} &
    \makecell[l]{AI, ML} &
    \makecell[l]{Full papers, strengths, weaknesses,\\questions, reviewer scores, \\rebuttal discussions,\\meta-reviews, editorial decisions,\\structured review reasoning traces} &
    \makecell[l]{13k+ samples} &
    \makecell[l]{Review generation, \\paper rating} \\

    \makecell[l]{Gen-Review\\ \citeyear{demetrio-gen-2025}} &
    \makecell[l]{OpenReview} &
    \makecell[l]{AI, ML} &
    \makecell[l]{Full papers, human reviews,\\LLM reviews } &
    \makecell[l]{32k+ papers,\\124k+ human reviews,\\81k+ LLM reviews} &
    \makecell[l]{Human--LLM review comparison,\\review bias analysis, \\AI-generated review detection} \\

    \makecell[l]{LimAgents\\ \citeyear{azher-limagents-2025}} &
    \makecell[l]{OpenReview, \\arXiv, \\ OpenAlex} &
    \makecell[l]{AI, ML} &
    \makecell[l]{Full papers,\\ limitations,\\ weaknesses} &
    \makecell[l]{2,700 papers,\\ 51k+ limitations} &
    \makecell[l]{Research limitation generation,\\ weakness identification, \\ pre-submission paper refinement} \\

    \makecell[l]{Reviews-STD\\ \citeyear{lu-agent-2025}} &
    \makecell[l]{Reviewer2~\citeyear{gao-reviewer2-2024}, \\SEA~\citeyear{yu-automated-2024}} &
    \makecell[l]{AI, ML} &
    \makecell[l]{Full papers, human reviews,\\meta-reviews, editorial decisions, \\strengths, weaknesses} &
    \makecell[l]{38k+ papers,\\144k+ reviews} &
    \makecell[l]{Review generation,\\editorial decision prediction, \\review quality assessment} \\

    \makecell[l]{RottenReviews\\ \citeyear{ebrahimi-rottenreviews-2025}} &
    \makecell[l]{OpenReview, \\F1000Research, \\SWJ, \\OpenAlex} &
    \makecell[l]{Multi} &
    \makecell[l]{Full papers, paper metadata, \\human reviews, \\reviewer profiles and expertise,\\editorial decisions, author responses,\\ review-quality annotations} &
    \makecell[l]{15k+ papers,\\55k+ reviews,\\9,532 reviewer profiles} &
    \makecell[l]{Review quality assessment,\\reviewer expertise analysis} \\

    \makecell[l]{TreeReview\\ \citeyear{chang-treereview-2025}} &
    \makecell[l]{SEA} &
    \makecell[l]{AI, ML} &
    \makecell[l]{Full papers,\\human reviews,\\actionable feedback comments} &
    \makecell[l]{80 papers} &
    \makecell[l]{Review generation,\\actionable feedback generation} \\

    \bottomrule

    \end{tabular}
    \end{adjustbox}
    
    \begin{tablenotes}
    \footnotesize
    \item \textit{Note:} AI = Artificial Intelligence; ML = Machine Learning; NLP = Natural Language Processing; SWJ = Semantic Web Journal.
    \end{tablenotes}

    \end{threeparttable}
    
\end{table}

Table~\ref{tab:multimodal-datasets} summarizes representative scholarly paper datasets of the second type, which include the full multimodal content of scholarly papers but lack review comments. These datasets are primarily developed for research on multimodal information understanding, supporting tasks such as figure understanding, figure classification, title generation, and code generation. With respect to data acquisition, most datasets obtain multimodal elements by directly cropping visual components, such as figures, tables, and formulas, from the original PDF files. In contrast, MMSCI~\citep{li-mmsci-graduate-2024} retrieves figures from official publisher websites, while Multimodal ArXiv~\citep{li-multimodal-2024} and SciGen~\citep{moosavi-scigen-2021} extract the images and tables from LaTeX source files. In terms of multimodal representation, the prevailing practice is to store visual elements, including figures, tables, and equations, as images in formats such as PNG or JPEG. A notable exception is SciGen~\citep{moosavi-scigen-2021}, which represents tables as linearized textual sequences and preserves their structural information through the use of special tokens. Although these datasets preserve multimodal paper content, the absence of review comments leaves them lacking direct supervision for modeling how human reviewers evaluate visual and structural elements. Supporting review-oriented tasks therefore requires pairing multimodal manuscripts directly with their corresponding review feedback.

\begin{table}[h]
    \centering
    \caption{Representative scholarly paper datasets containing multimodal content without review comments}
    \label{tab:multimodal-datasets}

    \setcellgapes{1.5pt}
    \makegapedcells
    \small
    
    \begin{threeparttable}

    \begin{adjustbox}{width=1.0\textwidth,keepaspectratio}

    \renewcommand{\arraystretch}{1.5} 

    \begin{tabular}{
        >{\raggedright\arraybackslash}p{0.09\textwidth}  
        >{\raggedright\arraybackslash}p{0.12\textwidth}  
        >{\raggedright\arraybackslash}p{0.07\textwidth}  
        >{\raggedright\arraybackslash}p{0.26\textwidth}  
        >{\raggedright\arraybackslash}p{0.12\textwidth}  
        >{\raggedright\arraybackslash}p{0.16\textwidth}  
        >{\raggedright\arraybackslash}p{0.18\textwidth} 
    }
    \toprule
    
    Name & Source & Domain & Content & Scale & Modality &  Application \\ \midrule
    
    \makecell[l]{MMAD\\ \citeyear{song-multidisciplinary-2025}} &
    \makecell[l]{Open-source\\ journal papers} &
    \makecell[l]{Multi} &
    \makecell[l]{Full contents, metadata, visual\\ information with captions and\\ descriptive context} &
    \makecell[l]{1.1m+ papers} &
    \makecell[l]{Images \\(figures, tables)} &
    \makecell[l]{Academic data\\ processing} \\
    
    \makecell[l]{MMSCI\\ \citeyear{li-mmsci-graduate-2024}} &
    \makecell[l]{Nature\\ Communications} &
    \makecell[l]{Multi} &
    \makecell[l]{Main body content, metadata,\\ figures with captions} &
    \makecell[l]{131k+ papers,\\ 742k+ figures} &
    \makecell[l]{Images (figures)} &
    \makecell[l]{Scientific article and\\ figure understanding} \\
    
    \makecell[l]{Sci-Cap+\\ \citeyear{yang-scicap-2024}} &
    \makecell[l]{Kaggle arXiv\\ dataset} &
    \makecell[l]{CS, ML} &
    \makecell[l]{Figures, OCR-extracted text and\\ bounding boxes, mention-\\paragraph pairs} &
    \makecell[l]{414k+ figures,\\ 12m+ words} &
    \makecell[l]{Images (figures)} &
    \makecell[l]{Figure caption\\ generation} \\
    
    \makecell[l]{ChartMimic\\ \citeyear{yang-chartmimic-2025}} &
    \makecell[l]{arXiv} &
    \makecell[l]{Multi} &
    \makecell[l]{Figure-code-instruction triplets} &
    \makecell[l]{4,800 triplets} &
    \makecell[l]{Images (figures)} &
    \makecell[l]{Visually grounded\\ code generation} \\
    
    \makecell[l]{Multimodal \\ArXiv\\ \citeyear{li-multimodal-2024}} &
    \makecell[l]{arXiv} &
    \makecell[l]{Multi} &
    \makecell[l]{Figures, captions, titles,\\ abstracts, multiple-choice\\ Q\&A pairs} &
    \makecell[l]{6.4m figures,\\ 3.9m captions,\\ 100k Q\&A pairs} &
    \makecell[l]{Images (figures)} &
    \makecell[l]{LVLM-based scientific\\ comprehension} \\
    
    \makecell[l]{ACL-FIG\\ \citeyear{karishma-acl-2023}} &
    \makecell[l]{ACL Anthology} &
    \makecell[l]{CL, NLP} &
    \makecell[l]{Figures, metadata} &
    \makecell[l]{151k+ tables,\\ 112k+ figures} &
    \makecell[l]{Images \\(figures, tables)} &
    \makecell[l]{Figure classification,\\ retrieval, generation} \\
    
    \makecell[l]{SciGen\\ \citeyear{moosavi-scigen-2021}} &
    \makecell[l]{arXiv} &
    \makecell[l]{CS} &
    \makecell[l]{Table-description pairs} &
    \makecell[l]{53k+ pairs} &
    \makecell[l]{Text with structural \\tokens (tables)} &
    \makecell[l]{Table description\\ generation} \\
    
    \bottomrule

    \end{tabular}
    \end{adjustbox}
    
    \begin{tablenotes}
    \footnotesize
    \item \textit{Note:} CS = Computer Science; ML = Machine Learning; CL = Computational Linguistics; NLP = Natural Language Processing;\\ LVLM = Large Vision-Language Model.
    \end{tablenotes}

  \end{threeparttable}
    
\end{table}

In contrast to the first two categories, datasets that jointly integrate multimodal scholarly paper content with review comments remain extremely scarce. In our review of existing resources, cPAPERS~\citep{sundar-cpapers-2024} emerged as the sole representative dataset in this category, underscoring the lack of benchmarks linking visual and structural paper elements with peer review feedback. As shown in table~\ref{tab:multimodal-comment-datasets}, cPAPERS focuses on conversational interactions by extracting question-answer pairs related to multimodal content from official reviews and comments of papers published at NeurIPS and ICLR on OpenReview, while simultaneously retrieving equations, figures, and tables from the corresponding arXiv LaTeX source files. In cPAPERS, tables and equations are represented in LaTeX format, whereas figures are stored as images. Designed to support the development of conversational assistants for scientific literature, cPAPERS aligns more closely with ASPR. However, it is organized around question-answer pairs associated with specific multimodal elements rather than complete manuscript-level review records. Effective ASPR modeling requires full manuscripts paired with complete reviewer reports that are precisely grounded in the specific manuscript versions on which those reports were based.  

\begin{table}[h]
    \centering
    \caption{Representative scholarly paper datasets containing multimodal content and review comments}
    \label{tab:multimodal-comment-datasets}

    \setcellgapes{1.5pt}
    \makegapedcells
    \small
    
    \begin{adjustbox}{width=1.0\textwidth,keepaspectratio}
    
    \renewcommand{\arraystretch}{1.5} 

    \begin{tabular}{
        >{\raggedright\arraybackslash}p{0.1\textwidth}  
        >{\raggedright\arraybackslash}p{0.1\textwidth}  
        >{\raggedright\arraybackslash}p{0.07\textwidth}  
        >{\raggedright\arraybackslash}p{0.22\textwidth}  
        >{\raggedright\arraybackslash}p{0.13\textwidth}  
        >{\raggedright\arraybackslash}p{0.20\textwidth}  
        >{\raggedright\arraybackslash}p{0.18\textwidth} 
    }
    \toprule

    Name & Source & Domain & Content & Scale & Modality &  Application \\ \midrule
    
    \makecell[l]{cPAPERS\\ \citeyear{sundar-cpapers-2024}} & 
    \makecell[l]{arXiv, \\OpenReview} & 
    \makecell[l]{Multi} & 
    \makecell[l]{Equations, tables, figures,\\ question-answer pairs,\\ contexts, references} & 
    \makecell[l]{2,350 papers,\\ 5,030 question-\\answer pairs} & 
    \makecell[l]{LaTeX (equations, tables); \\images (figures)} &  
    \makecell[l]{Conversational\\ assistant development} \\
    
    \bottomrule

    \end{tabular}
    \end{adjustbox}
\end{table}

Despite measurable progress driven by existing datasets, three structural limitations continue to constrain empirical advancement in ASPR research. First, domain coverage remains severely imbalanced. Most publicly available datasets concentrate on computer science and are predominantly derived from conferences hosted on platforms such as OpenReview, including NeurIPS and ICLR, with other disciplines substantially underrepresented. This disciplinary skew introduces systematic distribution bias, weakens cross-domain generalization, and restricts the development of ASPR systems capable of operating reliably across heterogeneous fields. Second, the alignment between review comments and manuscript versions is frequently compromised. Most existing datasets typically release only the final revised manuscripts, whereas review comments were written in response to an earlier submission. Because the original draft is often unavailable, or cannot be reliably aligned with the released version, reconstructing the authentic review context becomes infeasible. This temporal and semantic misalignment produces a structural inconsistency between the review comments and their intended targets. Third, datasets that systematically integrate multimodal manuscript content with review comments remain scarce. Most publicly available resources offer either multimodal paper content or review text in isolation, with comprehensive datasets bridging modalities remaining remarkably rare. Even within the few hybrid resources that exist, complete pairing between full manuscripts and complete reviewer reports is seldom available. This modality gap severely constrains the systematic development and rigorous benchmarking of multimodal ASPR systems. Taken together, these limitations impair the faithful modeling of real-world peer review processes and, in turn, diminish the practical viability of multimodal LLMs in academic evaluation.

\section{Dataset}
\label{sec:dataset}

This section details the data source, construction pipeline, and key properties of FMMD, a multimodal multidisciplinary dataset of open peer reviews from F1000Research.

\subsection{Data source}

F1000Research\footnote{https://f1000research.com/} is an open-access, multidisciplinary publishing platform designed to support rapid and transparent dissemination of scientific research. It publishes a wide range of research outputs, including research articles, data notes, software tools, and methodological articles, under a continuous publication model with fully open peer review and publicly accessible source data. Unlike traditional journals that conduct peer review prior to publication, F1000Research employs a post-publication, open peer review model. Manuscripts are published online shortly after submission, and peer review then proceeds transparently on the article page, with reviewer reports and author responses openly accessible. This model fosters transparency and supports sustained scholarly dialogue.

F1000Research categorizes its published articles into six core subject areas: Natural Sciences, Medical and Health Sciences, Social Sciences, Engineering and Technology, Agricultural and Veterinary Sciences, and Humanities and the Arts. We adopted this platform-defined taxonomy to characterize the disciplinary coverage of FMMD, extracting article-subject associations directly from the subject-area listings on the F1000Research platform. Because an article may belong to multiple subject areas, these associations also allow us to identify multidisciplinary articles and analyze their disciplinary distribution.

F1000Research adopts a versioned publication model in which each manuscript is released online shortly after submission as an initial version (Version 1). Authors may subsequently revise and update the manuscript to produce newer versions (e.g., Version 2, Version 3), with all versions remaining publicly accessible and independently citable. Each manuscript version is presented on a dedicated, structured HTML article page that integrates the complete scholarly content of the article with platform-maintained publication and peer review information. 

\begin{figure}[htb]
    \centering
    \includegraphics[width=\textwidth]{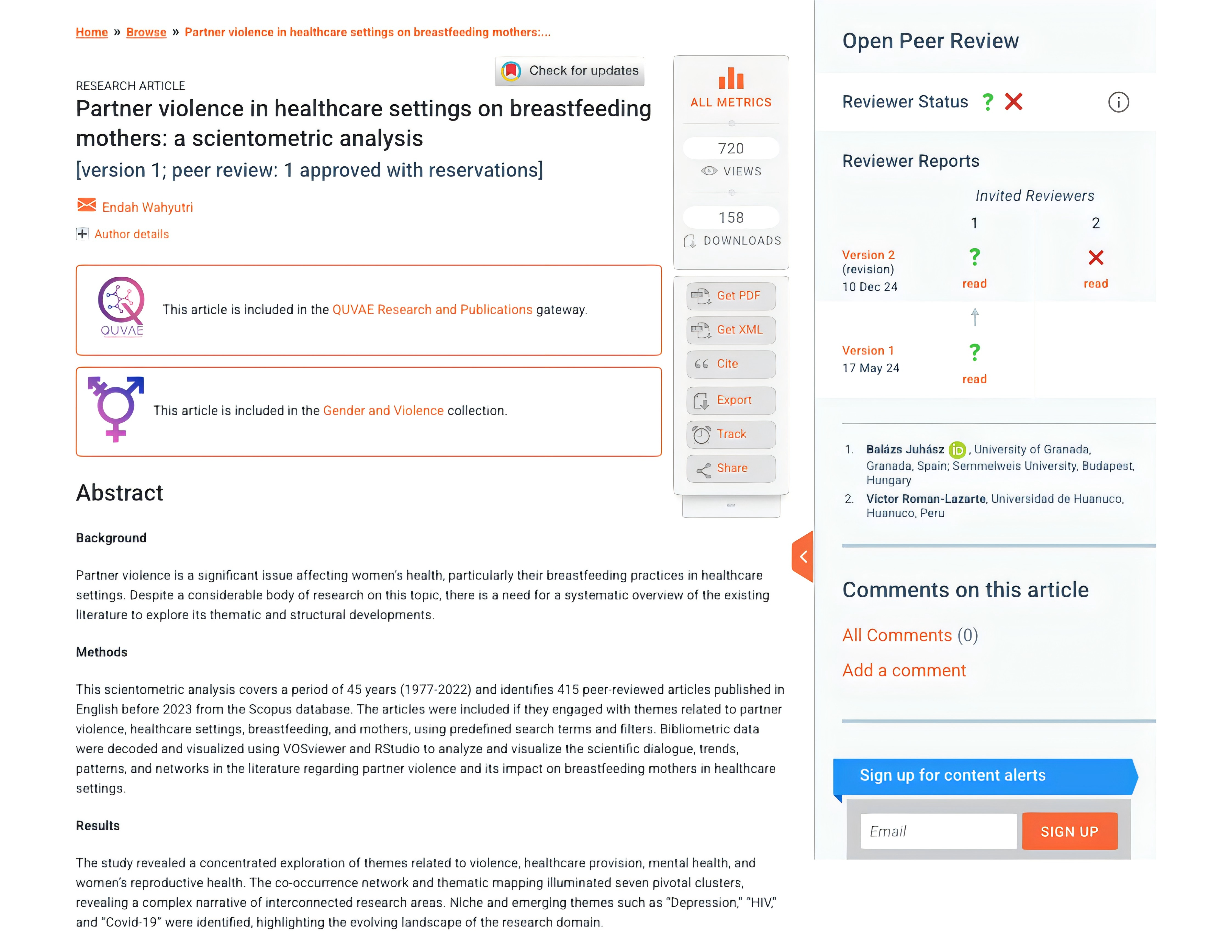}
    \caption{An example Version 1 article page on F1000Research}
    \label{fig:article-structure}
\end{figure}

Fig.~\ref{fig:article-structure} presents the HTML article page for Version 1 of a representative article. The central area displays core academic content, including the article title, author list, abstract, and manuscript body, flanked by two platform-level functional modules: Collections and Gateways. The Collections module groups articles into curated thematic collections to facilitate topic-based literature discovery, while the Gateways module assigns institutional or organizational portal labels, enabling the aggregation and presentation of research outputs at the institutional level through author affiliations.

\begin{figure}[htb]
    \centering

    \begin{minipage}[c]{0.495\textwidth}
        \raggedright
    
        \begin{subfigure}[c]{0.63\linewidth}
            \centering
            \includegraphics[width=\linewidth]{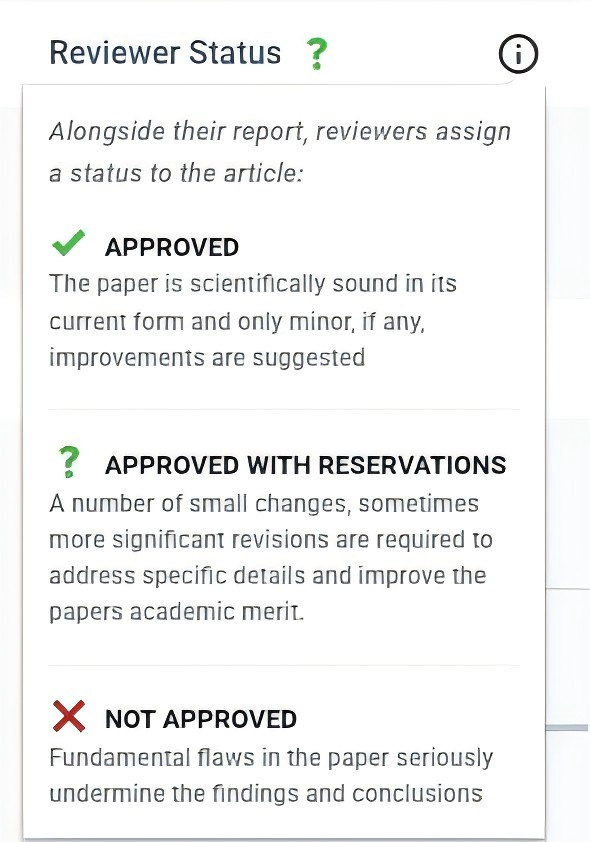}
            \caption{}
            \label{fig:three-status}
        \end{subfigure}
    
        \vspace{0.6cm}
    
        \begin{subfigure}[c]{1\linewidth}
            \centering
            \includegraphics[width=\linewidth]{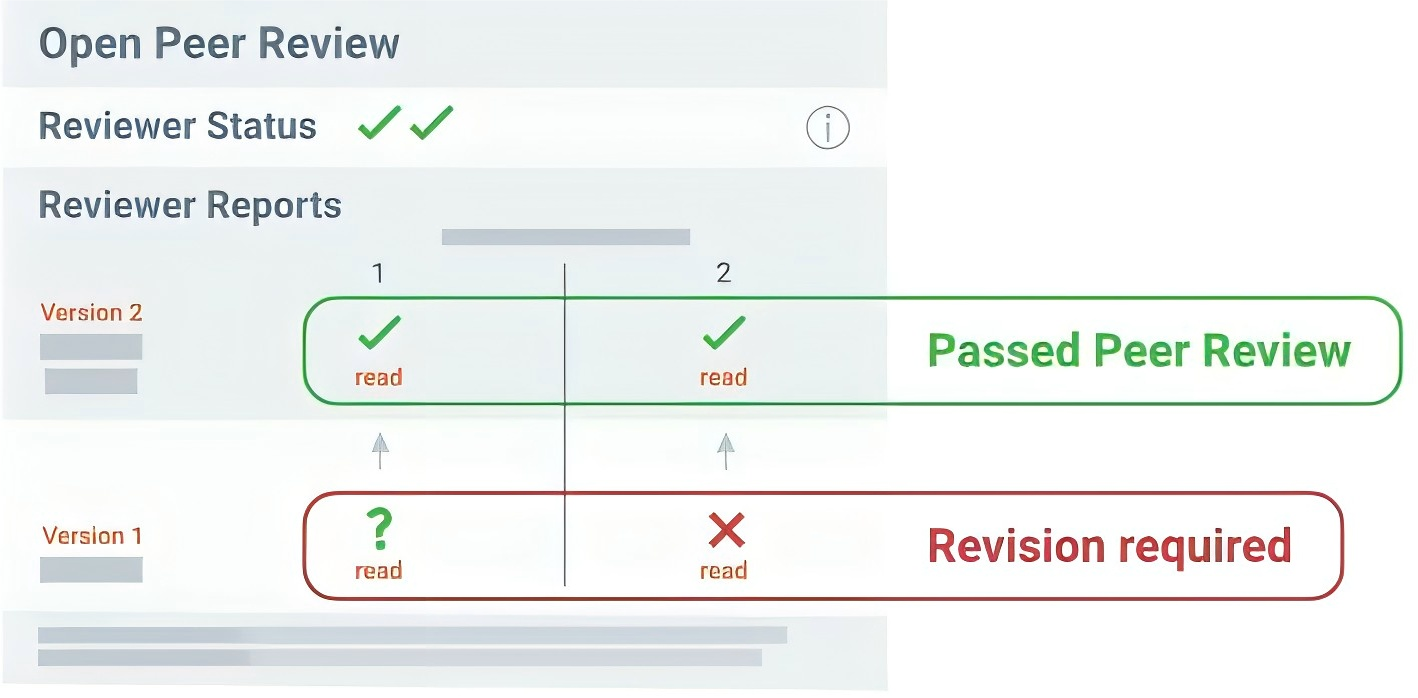}
            \caption{}
            \label{fig:passing-status}
        \end{subfigure}
    
    \end{minipage}
    \hspace{0.02\textwidth}
    \begin{subfigure}[c]{0.357\textwidth}
        \centering
        \includegraphics[width=\linewidth]{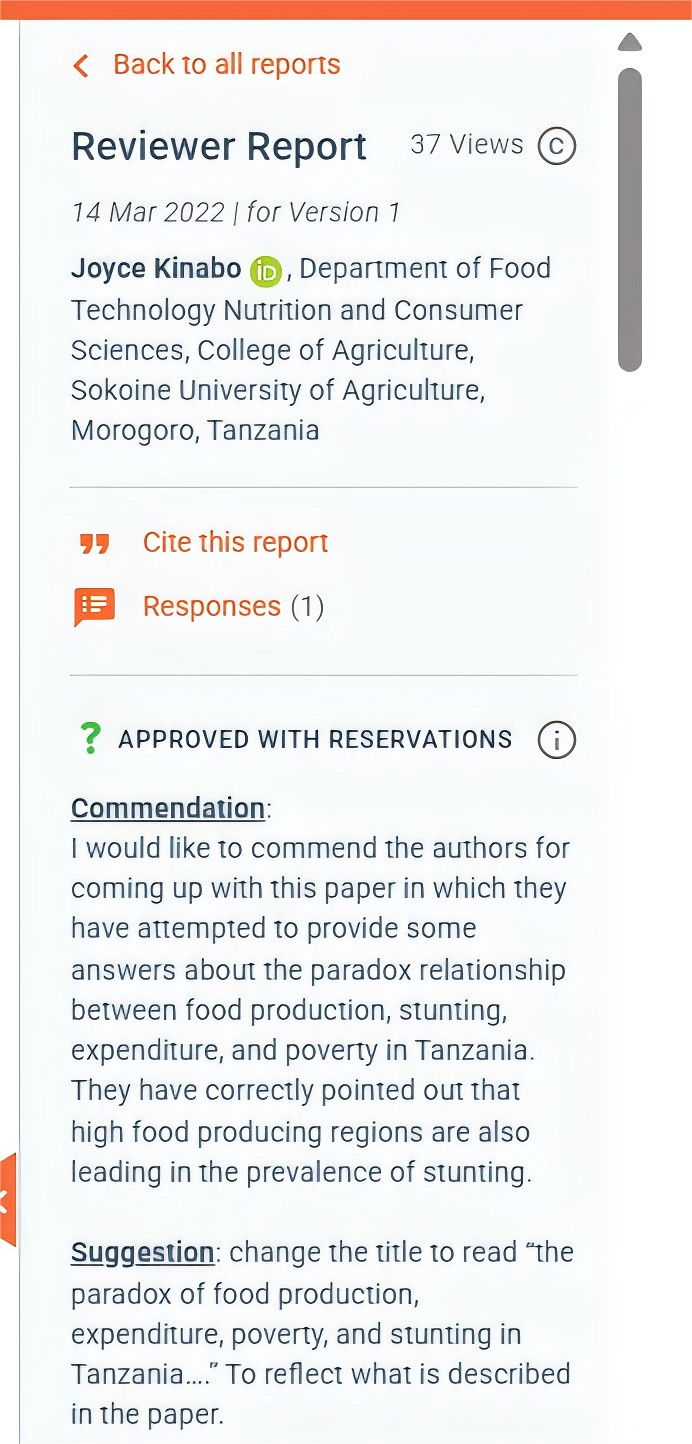}
        \caption{}
        \label{fig:reviewer-report-page}
    \end{subfigure}

    \caption{Peer review process on F1000Research: 
    (a) review status definitions, 
    (b) version progression and passing criteria, and 
    (c) an example Reviewer Report page.}
    \label{fig:peer-review-example}

\end{figure}

The right-hand side of the article page hosts the Open Peer Review module, which displays the review status and reviewer reports associated with each version. Peer-review evaluations are visually indicated by explicit status icons: a green checkmark for \textit{Approved}, a green question mark for \textit{Approved with Reservations}, and a red cross for \textit{Not Approved}, as illustrated in Fig.~\ref{fig:peer-review-example}\subref{fig:three-status}. An article version successfully passes peer review when it accumulates either two \textit{Approved} status icons, or one \textit{Approved} status icon together with two \textit{Approved with Reservations} status icons. Fig.~\ref{fig:peer-review-example}\subref{fig:passing-status} illustrates this versioned progression, where Version 1, marked as ``Revision required'', evolves into Version 2 that achieves ``Passed Peer Review''. Achieving this status determines the article's eligibility for indexing in bibliographic databases. Readers can click the ``read'' link beneath a reviewer entry to access the corresponding Reviewer Report page, which contains full review comments and author responses, as shown in Fig.~\ref{fig:reviewer-report-page}. Together, this versioned publication model and open peer review process provide a structured, transparent, and multi-turn record of manuscript evolution and evaluation, serving as a rich foundation for constructing FMMD.

\subsection{Construction pipeline}

This subsection details the three-stage construction pipeline for FMMD, where each sequential phase yields a progressively refined dataset.

In the first stage, we conducted article crawling and version retrieval on the F1000Research platform, collecting all articles published up to January 1, 2026. A total of 10,914 articles were indexed, and the complete HTML source pages for all available versions of each article were collected. To ensure data quality, articles lacking valid peer review information, specifically those without any reviewer reports, were excluded, as they did not contain analyzable peer review data. This filtering process resulted in a final set of 8,939 articles. For each retained article version, all referenced image resources were downloaded locally, and the corresponding remote URLs in the HTML source were replaced with local relative paths. This procedure preserves the original page layout and ensures a stable correspondence between textual content and visual elements in a local environment. The output of this stage comprises raw multimodal HTML files accompanied by their locally linked images.

In the second stage, we performed structured HTML parsing over all archived article versions to extract core metadata related to the open peer review process. The extracted information includes article identifiers, version numbers, and detailed peer review data for each version. Specifically, the article identifiers cover article ID, title, and version-specific URLs, while the peer review data comprise reviewer information, review status, full review comments, and corresponding author responses. All extracted metadata was consolidated into structured CSV files.

In the third stage, we conducted semantic analysis targeting multimodal content within peer review comments and author responses. Using the DeepSeek-V3.2 API in reasoner mode with the prompt provided in~\ref{subsubsec:app-multimodal-content-identification}, we performed context-aware identification of sentences that explicitly or implicitly refer to multimodal elements, including figures, tables, visual layouts, and other non-textual or structural components requiring visual interpretation. The identified sentences were subsequently grouped based on their semantic similarity, yielding a structured CSV dataset containing the results of the multimodal semantic analysis.

Across the three stages, the construction pipeline produces three complementary data products of multimodal HTML files with locally linked images, structured peer review metadata in CSV format, and multimodal semantic analysis results in CSV format. Together, these data products constitute the multimodal peer review dataset of FMMD used in this study. 

Following our previous work of Multidisciplinary Open Peer Review Dataset (MOPRD)~\citep{lin-moprd-2023}, FMMD is released in two formats, with the Native version containing the original HTML files, and the Processed version comprising carefully refined and structured data. The dataset is publicly available through our website.\footnote{http://www.linjialiang.net/publications/fmmd}

\subsection{Properties}

FMMD comprises 8,939 articles spanning multiple disciplines. Table~\ref{tab:statistical-features} provides a comprehensive statistical summary of the dataset across article-level metadata, reviewer reports, author responses, and multimodal elements.

\begin{table}[htb]
    \centering
    \caption{Statistical summary of the FMMD dataset}
    \label{tab:statistical-features}

    \setcellgapes{2pt}
    \makegapedcells
    \small

    \begin{adjustbox}{width=0.7\textwidth,keepaspectratio}

	\begin{tabular}{l l r}
	\toprule
        Category & Statistic & Value \\
        \midrule
        Articles
        & Total articles & 8,939 \\
        & Articles passing peer review & 5,755 \\
        & Revision required articles & 1,227 \\
        & Partially reviewed articles & 1,957 \\
        & Mean versions per article & 1.6 \\
        & Mean word count per article & 3,905 \\
        \midrule
        Reviewer reports
        & Total reviewer reports & 26,123 \\
        & Mean reviewer reports per article & 3.0 \\
        & Mean reviewer reports per version & 1.8 \\
        & Mean word count per review comment & 288.2 \\
        \midrule
        Author responses
        & Total author responses & 8,887 \\
        & Mean author responses per article & 1.0 \\
        & Mean author responses per version & 0.6 \\
        & Mean word count per author response & 30.5 \\
        \midrule
        Multimodal items
        & Total multimodal items & 49,082 \\
        & Mean multimodal items per version & 3.4 \\
        & Total multimodal-related review comments & 6,913 \\
        & Total multimodal-related author responses & 1,890 \\
        & Mean multimodal-related review comments per article & 0.8 \\
        & Mean multimodal-related review comments per version & 0.5 \\
        & Mean multimodal-related author responses per article & 0.2 \\
        & Mean multimodal-related author responses per version & 0.1 \\
        \bottomrule
        
    \end{tabular}
    \end{adjustbox}
\end{table}

\begin{figure}[htbp]
    \centering

    \begin{subfigure}[b]{0.531\textwidth}
        \centering
        \includegraphics[width=\textwidth]{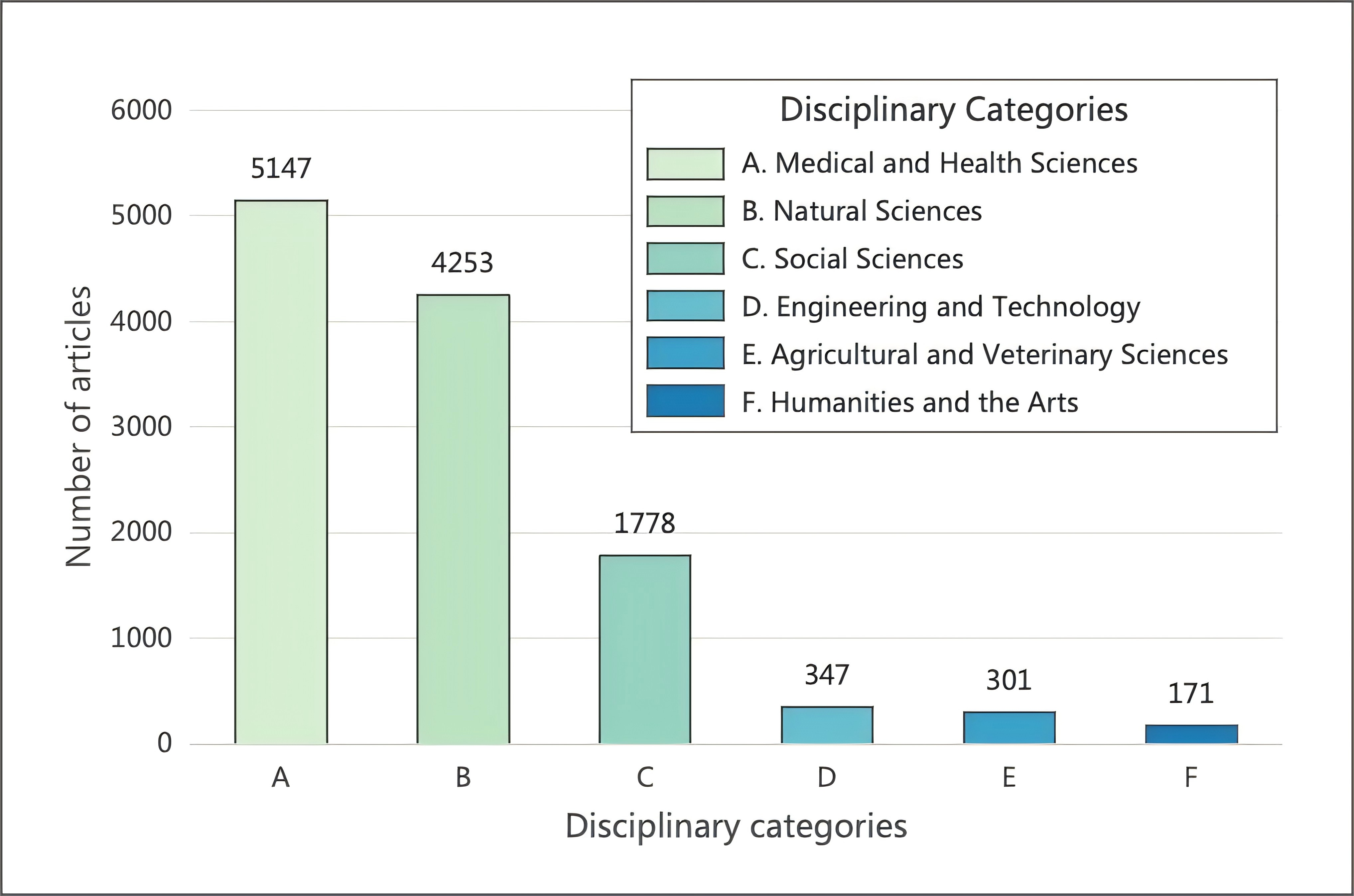}
        \caption{}
        \label{fig:subject-areas}
    \end{subfigure}
    \hfill
    \begin{subfigure}[b]{0.463\textwidth}
        \centering
        \includegraphics[width=\textwidth]{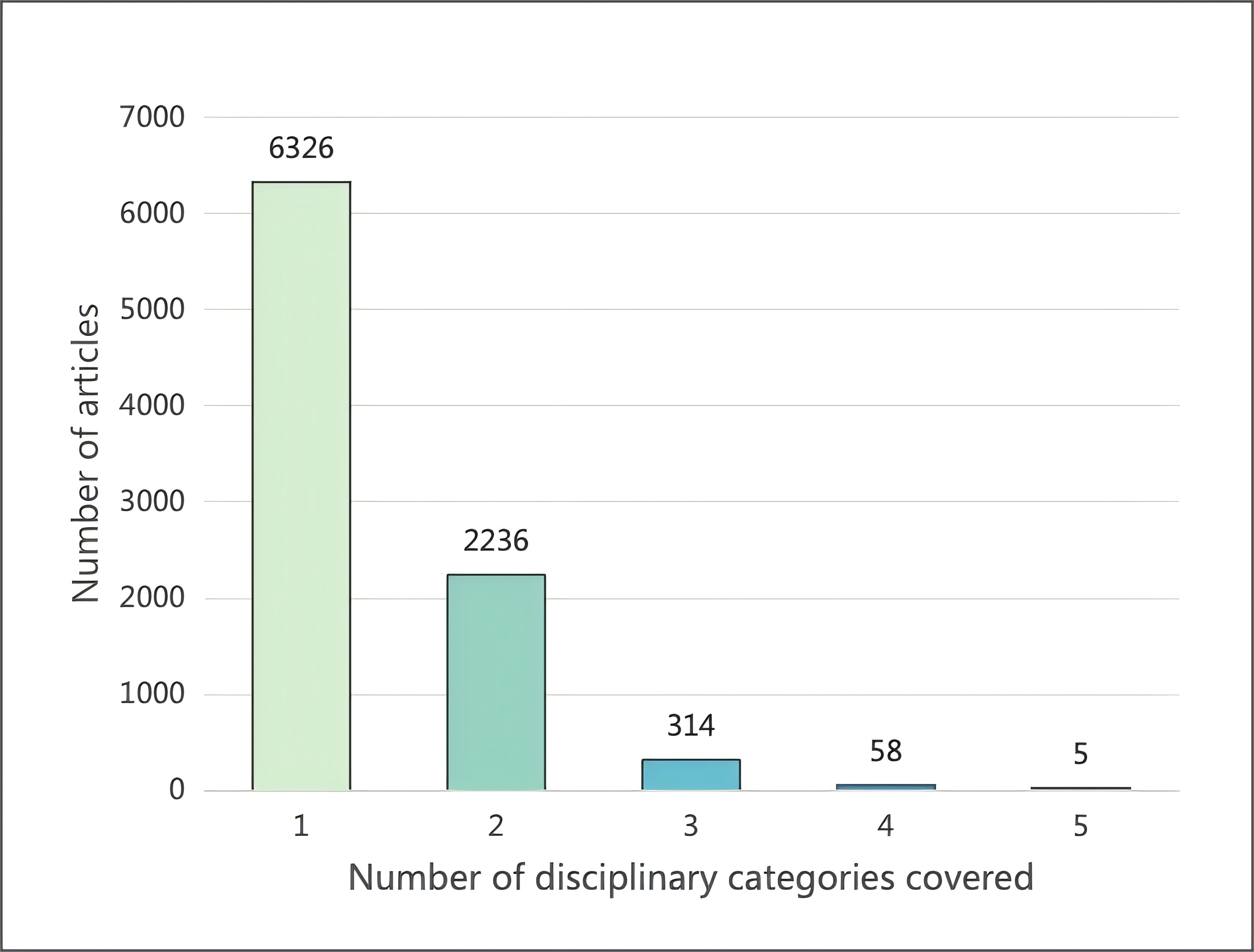}
        \caption{}
        \label{fig:disciplinary-category}
    \end{subfigure}

    \caption{Disciplinary distribution of FMMD: 
    (a) distribution of articles across disciplinary categories, and 
    (b) distribution of the number of disciplinary categories covered per article.}
    \label{fig:disciplinary-distribution}
\end{figure}

Fig.~\ref{fig:disciplinary-distribution}\subref{fig:subject-areas} illustrates the disciplinary breadth of the FMMD dataset. The distribution across disciplinary categories is characterized by a primary orientation toward Medical and Health Sciences and Natural Sciences, followed by a secondary cluster in Social Sciences. In contrast, Engineering and Technology, Agricultural and Veterinary Sciences, and Humanities and the Arts are comparatively underrepresented. This distribution provides a diverse benchmark for evaluating the domain-specific reasoning of ASPR systems beyond the Engineering and Technology sectors that dominate existing datasets. Furthermore, the aggregate category frequency surpasses the total article count because some articles are mapped to multiple disciplinary categories, reflecting the inherent interdisciplinary nature of modern scientific publication. However, as shown in Fig.~\ref{fig:disciplinary-distribution}\subref{fig:disciplinary-category}, the distribution of disciplinary coverage per article is strongly skewed toward single-discipline research. Specifically, 70.8\% of articles are confined to a single disciplinary category, while only 4.2\% span three or more disciplines. This steep decline indicates that, while limited cross-disciplinary engagement exists within FMMD, highly integrative interdisciplinarity remains rare.

\section{Experiments}
\label{sec:experiments}

To evaluate the utility of FMMD for multimodal scientific reasoning in ASPR, we formulate three benchmark tasks: (1) visual-semantic consistency classification, which assesses the alignment between visual content and textual descriptions; (2) figure-related review comment generation, which examines the generation of critical and constructive feedback on visual elements; and (3) editorial decision prediction, which investigates holistic judgment by integrating textual and visual information from the full manuscript.

We adopt Qwen3-VL-8B-Instruct as the primary baseline model, given its strong general-purpose vision-language capabilities, open-source availability, and suitability for fine-tuning. To rigorously evaluate the empirical utility of FMMD, we compare the base model with its FMMD-fine-tuned counterpart under controlled experimental settings.

For each task, the corresponding experimental dataset was partitioned into training, validation, and test sets in an 8:1:1 ratio. The models were fine-tuned using task-specific objectives, including binary classification and autoregressive generation. All experiments were conducted on the AutoDL computing platform.\footnote{https://www.autodl.com/}

\subsection{Visual-semantic consistency classification}

Visual elements, such as figures, are essential for conveying experimental findings and methodological information in scholarly papers. Consequently, assessing the consistency between visual content and textual descriptions is an important aspect of peer review. This experiment evaluates whether figure-related feedback in FMMD provides effective domain-specific supervisory signals for identifying visual-semantic inconsistencies.

\textbf{Task definition.}
We formulate this task as a binary classification problem. Given a figure $Fig$, its title and caption combined into a single textual field $Cap$, and the surrounding contextual text $Ctx$, the model predicts a label $y \in {0, 1}$, where $y=1$ denotes a visual-semantic inconsistency and $y=0$ denotes consistency.

\textbf{Data construction.}

We utilized the DeepSeek-V4-Pro-0813\footnote{The data for Task (1) was reparsed using the updated DeepSeek-V4-Pro-0813 API available at the time of execution instead of the initial DeepSeek-V3.2 API in reasoner mode.} and MiniMax-M2.7-highspeed APIs as independent annotators, using the prompt provided in~\ref{subsubsec:app-exp1-review-comment-extraction}, to parse a total of 26,123 reviewer reports for explicit mentions of visual-semantic inconsistencies. To improve annotation reliability, we retained only the samples that both DeepSeek and MiniMax independently classified as visual-semantic inconsistencies. This process ultimately yielded 1,488 positive samples (y=1). Negative samples (y=0) were constructed by randomly selecting figures from articles with a Passed Peer Review status for which both models identified no figure-related issues in their review records. To prevent sampling bias, positive and negative samples were balanced at a 1:1 ratio across the training, validation, and test sets.

\textbf{Experimental design.}

To investigate the impact of FMMD-based fine-tuning and context length on visual-semantic consistency classification, we adopted a controlled 2×2 factorial design. The four experimental conditions consisted of a base model with sentence-level context, a base model with paragraph-level context, an FMMD fine-tuned model with sentence-level context, and an FMMD fine-tuned model with paragraph-level context. To ensure robust evaluation, all experiments were conducted twice independently using different data partitions, and the reported results represent the average performance across the two runs. The prompt used for model fine-tuning and evaluation is provided in~\ref{subsubsec:app-exp1-task-instruction}.

\textbf{Results analysis.}
Table~\ref{tab:performance-generation} presents model performance across the four experimental conditions on the visual-semantic consistency classification task. The results demonstrate that FMMD fine-tuning consistently yields performance gains across all evaluation metrics under both contextual settings. This indicates that supervisory signals provided by open peer review data effectively enhance a model's ability to identify visual-semantic inconsistencies.

Specifically, fine-tuning on FMMD drives a substantial surge in Recall across both contextual settings compared to other metrics, with relative increases of 43.0\% for sentence-level context and 66.9\% for paragraph-level context. This marked gain indicates that the supervisory signals provided by FMMD help the model detect visual-semantic inconsistencies more effectively. In ASPR scenarios, prioritizing high Recall is particularly critical because failing to flag a genuine visual-semantic inconsistency carries a significantly higher cost than generating a small number of false-positive alerts.

Regarding input representation, sentence-level context outperforms paragraph-level context on most evaluation metrics across both base and fine-tuned models. This indicates that in visual-semantic consistency classification, expanding the contextual length does not guarantee performance improvements. Instead, the additional contextual information may introduce irrelevant content that distracts the model from the relationship between the figure and its textual context.

Overall, these findings validate the effectiveness of FMMD in visual-semantic consistency classification and demonstrate that figure-text feedback accumulated during open peer review provides valuable domain-specific supervisory signals for visual-semantic inconsistencies.

\begin{table}[htb]
    \centering
    \caption{Visual-semantic consistency classification}
    \label{tab:performance-consistency}

    \setcellgapes{2pt}
    \makegapedcells
    \small

    \begin{adjustbox}{width=0.63\textwidth,keepaspectratio}

	\begin{tabular}{l l r r r r}
	\toprule
	Fine-tuning & Input context & Accuracy & Precision & Recall & F1-score \\
        \midrule
         None 
                & Sentence & 0.537  & 0.556  & 0.408  & 0.470  \\
                & Paragraph & 0.505  & 0.506  & 0.360  & 0.420  \\
        FMMD
                & Sentence & \textbf{0.596} & \textbf{0.611} & 0.584  & \textbf{0.594} \\
                & Paragraph & 0.580  & 0.577  & \textbf{0.600}  & 0.588 \\
        \bottomrule
        
    \end{tabular}
    \end{adjustbox}
\end{table}

\subsection{Figure-related review comment generation}

Beyond identifying inconsistencies, high-quality peer review requires actionable, detailed, and constructive critiques of the clarity, scientific rigor, and presentation quality of visual elements. This experiment evaluates whether fine-tuning on FMMD facilitates the generation of figure-related review comments that align with critical feedback patterns in authentic peer review.

\textbf{Task definition.}
We formulate this task as a multimodal conditional generation problem. Given a figure $Fig$, its title and caption combined into a single textual field $Cap$, and the surrounding contextual text $Ctx$, the model generates a natural language review comment $Cmt$ that critically evaluates the visual element.

\textbf{Data construction.}

We filtered a total of 26,123 reviewer reports using the DeepSeek-V3.2 API in reasoner mode and the GLM-5-Turbo API as independent annotators, using the prompt provided in~\ref{subsubsec:app-exp2-review-comment-extraction}, to identify comments containing negative, figure-related feedback. A comment was considered figure-related if it referenced a specific figure identifier (e.g., Fig. 1) and raised a specific concern regarding the figure. For each identified comment, we extracted the corresponding figure, along with its title, caption, and the surrounding contextual text from its original location in the article. To ensure consistency, when multiple comments referred to the same figure, they were merged into a single consolidated target sequence. This process yielded a dataset of 4,088 figure-review comment pairs.

\textbf{Experimental design.} 
To investigate the impact of FMMD-based fine-tuning and context length on figure-related review comment generation, we adopted a controlled 2×2 factorial design. The four experimental conditions consisted of a base model with sentence-level context, a base model with paragraph-level context, an FMMD fine-tuned model with sentence-level context, and an FMMD fine-tuned model with paragraph-level context. To ensure robust evaluation, all experiments were conducted twice independently using different random seeds, and the reported results represent the average performance across the two runs. The prompt used for model fine-tuning and evaluation is provided in~\ref{subsubsec:app-exp2-task-instruction}.

\textbf{Results analysis.}
Table~\ref{tab:performance-generation} presents model performance across the four experimental conditions on the figure-related review comment generation task. The results demonstrate that FMMD fine-tuning yields substantial improvements in surface-level lexical overlap and semantic alignment, highlighting its advantage in capturing domain-specific critique patterns. Conversely, BARTScore remains near baseline levels.

Specifically, fine-tuning on FMMD drives substantial gains across all ROUGE variants. The sharp increase in ROUGE-2 suggests that FMMD helps the model capture domain-specific terminology and multi-word critique patterns rather than relying primarily on generic evaluation templates. For semantic alignment, although the improvement in BERTScore was modest, it increased consistently following FMMD fine-tuning. This suggests that fine-tuning improves semantic similarity in addition to increasing lexical overlap.

Regarding input representation, context length exerts a relatively limited influence on task performance. Across both the baseline and FMMD-fine-tuned models, expanding context from the sentence level to the paragraph level yields only marginal gains. This observation indicates that generating figure-related comments depends primarily on the visual content itself and its immediately adjacent textual descriptions.

Overall, these findings confirm that FMMD fine-tuning enhances the model's ability to capture domain-specific lexical and semantic patterns, enabling the generation of figure critiques that more closely align with authentic peer reviews.

\begin{table}[htb]
    \centering
    \caption{Figure-related review comment generation}
    \label{tab:performance-generation}

    \setcellgapes{2pt}
    \makegapedcells
    \small

    \begin{adjustbox}{width=0.81\textwidth,keepaspectratio}

	\begin{tabular}{l l r r r r r}
	\toprule
	Fine-tuning & Input context & ROUGE-1 & ROUGE-2 & ROUGE-L & BERTScore & BARTScore \\
        \midrule
        None 
                & Sentence & 10.865 & 1.535 & 9.741 & 80.787 & -4.588 \\
                & Paragraph & 11.367 & 1.341 & 10.161 & 81.286  & \textbf{-4.582} \\
        FMMD
                & Sentence & 16.322  & \textbf{7.329} & 15.467  & 81.126  & -4.679  \\
                & Paragraph & \textbf{18.280} & 6.377  & \textbf{16.716} & \textbf{82.441} & -4.626  \\
        \bottomrule
        
    \end{tabular}
    \end{adjustbox}
\end{table}

\subsection{Editorial decision prediction}
Editorial decisions in peer review require a holistic assessment of textual arguments and visual evidence across the entire manuscript. This experiment evaluates whether FMMD improves editorial decision prediction and whether incorporating multimodal visual information provides additional value over text-only representations.

\textbf{Task definition.}
We formulate this task as a binary classification problem. Given the full-text HTML of an article version $T_{full}$ and its set of all embedded figures $\{Fig_1, Fig_2, ..., Fig_n\}$, the model predicts the final decision $y \in \{0, 1\}$, where $y=1$ denotes Approved and $y=0$ denotes Not Approved.

\textbf{Data construction.}
We constructed positive samples (y=1) by selecting articles in their final version for which all three reviewers assigned an Approved status. This yielded an initial pool of 1,629 positive samples. Conversely, negative samples were drawn from Version 1 articles for which all three reviewers assigned a Not Approved status, totaling 639 negative samples (y=0). We randomly undersampled the positive class to obtain a balanced dataset of 1,278 samples, with 639 samples in each class. Each data instance comprised the full-text HTML document, all embedded figures, and the corresponding editorial decision status.

\textbf{Experimental design.} 
To investigate the impact of FMMD fine-tuning and input modalities on editorial decision prediction, we adopted a controlled 2×2 factorial design. The four experimental conditions  consisted of a base model with text-only input, a base model with multimodal input, an FMMD fine-tuned model with text-only input, and an FMMD fine-tuned model with multimodal input. To ensure robust evaluation, all experiments were conducted twice independently using different data partitions, and the reported results represent the average performance across the two runs. The prompt used for model fine-tuning and evaluation is provided in~\ref{subsubsec:app-exp3-task-instruction}.

\textbf{Results analysis.}
Table~\ref{tab:performance-prediction} presents the experimental results for the editorial decision prediction task. The results demonstrate that both multimodal input and FMMD fine-tuning independently and in combination boost predictive performance, with their combination achieving peak results across all metrics.

Specifically, the supervisory benefits introduced by FMMD fine-tuning were equally pronounced in decision modeling. Across both text-only and multimodal settings, FMMD fine-tuning substantially elevates overall metrics, exhibiting a particularly strong impact on Recall with relative increases of 58.6\% and 34.6\%, respectively. This demonstrates that the supervisory signals provided by FMMD help the model better leverage manuscript-level information, thereby improving decision prediction performance.

Regarding input representation, incorporating visual elements consistently outperforms text-only baselines under identical fine-tuning setups. Compared to using text-only input, introducing visual context yields relative improvements exceeding 20\% in both Recall and F1-score on the base model. This performance gap reveals that visual figures contain critical evaluative signals that textual content alone fails to convey, underscoring the value of visual-textual fusion in holistic paper evaluation.

Overall, these findings validate the utility of FMMD in high-level editorial decision modeling and confirm that preserving multimodal manuscript content provides a far more complete basis for ASPR research.

\begin{table}[htb]
    \centering
    \caption{Editorial decision prediction}
    \label{tab:performance-prediction}

    \setcellgapes{2pt}
    \makegapedcells
    \small

    \begin{adjustbox}{width=0.63\textwidth,keepaspectratio}

	\begin{tabular}{l l r r r r}
	\toprule
	Fine-tuning & Input & Accuracy & Precision & Recall & F1-score \\
        \midrule
        None 
                & Text-only & 0.570 & 0.582 & 0.453 & 0.502 \\
                & Multimodal & 0.629  & 0.634  & 0.609  & 0.621 \\
        FMMD
                & Text-only & 0.680 & 0.667 & 0.719 & 0.692 \\
                & Multimodal & \textbf{0.746} & \textbf{0.716} & \textbf{0.820}  & \textbf{0.764}  \\
        \bottomrule
        
    \end{tabular}
    \end{adjustbox}
\end{table}

\section{Conclusion}
\label{sec:concl}

In this work, we present FMMD, a large-scale, multimodal, and multidisciplinary open peer review dataset built upon F1000Research that addresses a critical gap in ASPR. To the best of our knowledge, FMMD is the first large-scale resource to capture the iterative and multimodal nature of the open peer review process across diverse disciplines. By integrating 8,939 articles alongside their full version histories, embedded visual elements, structured HTML representations, versioned reviewer reports, and corresponding author responses, the dataset enables systematic investigation of peer review as a multimodal and iterative evaluation process. Our empirical evaluations demonstrate that fine-tuning with FMMD significantly improves performance across three core evaluation tasks of visual-semantic consistency classification, figure-related review comment generation, and editorial decision prediction. Crucially, our findings underscore the important role of multimodal information in the ASPR process. As multimodal LLMs continue to evolve, FMMD positions itself as a key enabler for advancing ASPR from predominantly text-based assessment toward comprehensive multimodal evaluation. By releasing FMMD publicly, we aim to facilitate future research toward more comprehensive, multimodal, and human-aligned approaches to ASPR.

\section*{Acknowledgments}
This work is funded by the Special Innovative Project of Regular Higher Education Institutions in Guangdong Province (2025KTSCX209), the Research Initiation Fund Project of Guangzhou Institute of Science and Technology (2023KYQ184), the Guangdong Province Key Construction Discipline Research Capacity Enhancement Project (2024ZDJS101), and the Key Area Special Project of Regular Higher Education Institutions in Guangdong Province (2025ZDZX3054). Special and heartfelt gratitude goes to the corresponding author's wife Fenmei Zhou, for her understanding and love. Her unwavering support and continuous encouragement enable this research to be possible.

\section*{Declaration of competing interest}
The authors declare that they have no known competing financial interests or personal relationships that could have appeared to influence the work reported in this paper.

\section*{Declaration of generative AI and AI-assisted technologies use}
During the preparation of this work, Gemini 3.6 Flash was applied to draft the initial version of Fig. 1. After engaging with this tool, the authors carefully reviewed and revised all AI-generated content to ensure factual accuracy, logical coherence, and academic rigor. The authors take full responsibility for the content of this publication.

\appendix

\section{Prompts used in dataset construction and experiments}
\label{sec:app-prompts}

For reproducibility and transparency, we provide the complete prompts used in the construction of FMMD and in the three downstream experiments. These prompts served two purposes: identifying and extracting task-specific information from peer review records, and providing task instructions for the downstream experimental tasks. In Experiments 1 and 2, separate prompts were used for sample extraction and task execution, whereas Experiment 3 required only a task prompt.

\subsection{Dataset construction}
\label{subsec:app-dataset-construction}

\subsubsection{Multimodal-content identification}
\label{subsubsec:app-multimodal-content-identification}

The following prompt was used to identify sentences in peer review comments and author responses that referred to multimodal content. The identification was performed through semantic and context-aware analysis rather than keyword matching.

\begin{lstlisting}[style=promptstyle]
I want you help me analyze the following peer review comments and author responses. 
My requirement is to extract only the sentences that refer to true multimodal content, not just keywords. A multimodal reference is valid ONLY when the reviewer is talking about visual, graphical, or layout-related information that requires non-textual interpretation.

You must apply a semantic, context-aware evaluation, not keyword matching. Output all the mulitimodel sentences, group them by semantic meaning.
\end{lstlisting}

\subsection{Visual-semantic consistency classification}
\label{subsec:app-visual-semantic-consistency-classification}

\subsubsection{Review comment extraction}
\label{subsubsec:app-exp1-review-comment-extraction}

The following prompt was used to identify and extract review comments describing inconsistencies between figures and their corresponding textual descriptions.

\begin{lstlisting}[style=promptstyle]
I need you to analyze the provided peer review comments to identify and extract the specific figure numbers and the reviewer's verbatim remarks regarding inconsistencies between the figures and the text.

[What constitutes an inconsistency?]
Please extract all comments that report a discrepancy meeting the following criteria:
- Data/Logical Conflicts: Contradictory numerical values or trends (e.g., the text states a value of 1, but the figure shows 3; the text describes an increase, while the figure displays a decrease).
- Missing/Inconsistent Descriptions: The text describes specific elements not visible in the figure, or the figure displays features that remain unexplained in the text.
- Label/Legend Conflicts: The text specifies a certain meaning for a label or color (e.g., "red represents Group A"), but the figure or its legend indicates otherwise.
- Context Mismatch: Experimental setups, sample sizes, or conditions mentioned in the text do not align with what is visually depicted in the figure.
- Citation Errors: The text cites a specific figure to support a conclusion, but the figure fails to provide that support or directly contradicts the claim.
General Discrepancies: Any instance where the reviewer explicitly states that the text and the figure are inconsistent or poorly aligned.

[Exclusion Criteria] (Exclude the following):
- Pure Quality Issues: Complaints such as "blurry" or "low resolution," unless this causes a specific inconsistency with the text.
- Pure Formatting Issues: Comments on font size, line thickness, etc., without reference to textual inconsistency.
- Isolated Figure Evaluation: Critiques focusing solely on the figure itself without addressing its relationship to the written description.

[Core Requirements]:
Focus: Identify inconsistencies between the text and the figure.
Verbatim Extraction: You must extract the original sentences verbatim.
Identification: For each identified issue, specify the relevant figure number (e.g., "Figure 1", "Figure A") as mentioned in the comment.

[Output Format]:
You must output the results in a raw JSON format as follows:
If issues are found: {"results": [{"fig_no": "Figure X", "reviewer_comment": "verbatim text..."}]}
If NO issues are found: {"results": []}

Only output the raw JSON. Do not include any conversational filler or explanations.
\end{lstlisting}

\subsubsection{Task instruction}
\label{subsubsec:app-exp1-task-instruction}

The following prompt was used as the task instruction for visual-semantic consistency classification.

\begin{lstlisting}[style=promptstyle]
Carefully examine the provided figure and its accompanying text description.Compare the visual content of the figure with the claims made in the description, and determine whether any inconsistency exists between them. 
Check whether the description misrepresents the visual content, contradicts the displayed data or trends, refers to nonexistent visual elements, contains mismatched labels or annotations, or makes unsupported claims that are not evident from the figure. 
If any discrepancy, contradiction, mismatch, or unsupported claim is identified, output 1. 
If the figure and description are fully accurate and semantically consistent, output 0. 
Output only a single digit (0 or 1) without any additional text, explanation, or punctuation.
\end{lstlisting}

\subsection{Figure-related review comment generation}
\label{subsec:app-figure-related-review-comment-generation}

\subsubsection{Review comment extraction}
\label{subsubsec:app-exp2-review-comment-extraction}

The following prompt was used to identify and extract negative reviewer feedback and revision suggestions related to figures and images.

\begin{lstlisting}[style=promptstyle]
I need you to analyze the peer review comments provided to extract negative feedback or revision suggestions specifically regarding figures and images.

[Extraction Logic]
- Extract ONLY: Sentences expressing dissatisfaction, doubt, suggestions for revision, or identification of errors (e.g., blurred labels, inappropriate coloring, logical inconsistencies, low resolution, problem between data and text, etc.).
- Exclude: Purely descriptive or complimentary sentences.
- Requirement: Extract the original sentences verbatim. No paraphrasing.

[Output Format]
Return strictly a JSON object. Do not include any introductory text, conversational fillers, or explanations. 
- If issues are found: {"results": [{"fig_id": "Figure X", "reviewer_comment": "..."}]}
- If NO issues are found: {"results": []}

Only output the raw JSON.
\end{lstlisting}

\subsubsection{Task instruction}
\label{subsubsec:app-exp2-task-instruction}

The following prompt was used as the task instruction for generating figure-related review comments.

\begin{lstlisting}[style=promptstyle]
Carefully examine the provided figure and its accompanying text description. Generate a rigorous, precise, and professional review comment evaluating their quality and accuracy based on the following two steps. 
First, check for issues within the figure itself: Identify any presentation or clarity flaws, such as illegible text (font size too small), spelling or typographical errors within labels, mathematical inaccuracies in charts, or poor visual resolution. 
Second, identify discrepancies between the figure and its description: Compare the data, trends, or visual evidence shown in the figure with the claims made in the text. 
State explicitly if the description misrepresents the visual content, contradicts the displayed data, or makes unsupported claims. 
If any issues are identified, describe them clearly in a formal, academic tone. If no issues are found, state that the figure and its description are acceptable. 
Output the review comment directly.
\end{lstlisting}

\subsection{Editorial decision prediction}
\label{subsec:app-editorial-decision-prediction}

\subsubsection{Task instruction}
\label{subsubsec:app-exp3-task-instruction}

The following prompt was used as the task instruction for editorial decision prediction.

\begin{lstlisting}[style=promptstyle]
Evaluate the provided manuscript based on F1000Research publication standards, incorporating both the manuscript's textual content and its associated figures. 
The assessment must focus on scientific soundness, methodological rigor, and the validity of the data. Determine if the manuscript is qualified for publication. 
Output only the label: 'Accepted' or 'Not Accepted'.
\end{lstlisting}

\bibliography{mybib}

\begin{thebibliography}{48}
\expandafter\ifx\csname natexlab\endcsname\relax\def\natexlab#1{#1}\fi
\providecommand{\url}[1]{\texttt{#1}}
\providecommand{\href}[2]{#2}
\providecommand{\path}[1]{#1}
\providecommand{\DOIprefix}{doi:}
\providecommand{\ArXivprefix}{arXiv:}
\providecommand{\URLprefix}{URL: }
\providecommand{\Pubmedprefix}{pmid:}
\providecommand{\doi}[1]{\href{http://dx.doi.org/#1}{\path{#1}}}
\providecommand{\Pubmed}[1]{\href{pmid:#1}{\path{#1}}}
\providecommand{\bibinfo}[2]{#2}
\ifx\xfnm\relax \def\xfnm[#1]{\unskip,\space#1}\fi
\bibitem[{AAAI(2025)}]{aaai-instructions-2025}
\bibinfo{author}{AAAI} (\bibinfo{year}{2025}).
\newblock \bibinfo{title}{{Instructions for AAAI-26 Reviewers - AAAI}}.
\newblock \URLprefix
  \url{https://aaai.org/conference/aaai/aaai-26/instructions-for-aaai-26-reviewers/}.
\bibitem[{ACM(2025)}]{acm-evaluating-2025}
\bibinfo{author}{ACM} (\bibinfo{year}{2025}).
\newblock \bibinfo{title}{{Evaluating the Paper}}.
\newblock \URLprefix
  \url{https://reviewers.acm.org/training-course/evaluating-the-paper}.
\bibitem[{Azher et~al.(2026)Azher, Guo \& Alhoori}]{azher-limagents-2025}
\bibinfo{author}{Azher, I.~A.}, \bibinfo{author}{Guo, Z.}, \&
  \bibinfo{author}{Alhoori, H.} (\bibinfo{year}{2026}).
\newblock \bibinfo{title}{{Multi-agent LLMs for generating research
  limitations}}.
\newblock {\it \bibinfo{journal}{arXiv preprint arXiv:2601.11578}\/}, .
\bibitem[{Cai et~al.(2024)Cai, Cai, Yang, Wang, Yao, Gao, Chang, Li, Xu \&
  Wang}]{cai-uni-2024}
\bibinfo{author}{Cai, H.}, \bibinfo{author}{Cai, X.}, \bibinfo{author}{Yang,
  S.}, \bibinfo{author}{Wang, J.}, \bibinfo{author}{Yao, L.},
  \bibinfo{author}{Gao, Z.}, \bibinfo{author}{Chang, J.}, \bibinfo{author}{Li,
  S.}, \bibinfo{author}{Xu, M.}, \& \bibinfo{author}{Wang, C.}
  (\bibinfo{year}{2024}).
\newblock \bibinfo{title}{{Uni-SMART: Universal science multimodal analysis and
  research transformer}}.
\newblock {\it \bibinfo{journal}{arXiv preprint arXiv:2403.10301}\/}, .
\bibitem[{{Cambridge University Press}(2024)}]{cambridge-core-2024}
\bibinfo{author}{{Cambridge University Press}} (\bibinfo{year}{2024}).
\newblock \bibinfo{title}{{Core editorial policies for journals}}.
\newblock \URLprefix
  \url{https://www.cambridge.org/core/services/publishing-ethics/core-editorial-policies-journals}.
\bibitem[{Chang et~al.(2025)Chang, Li, Zhang, Kong, Wu, So, Guo, Zhu \&
  Wong}]{chang-treereview-2025}
\bibinfo{author}{Chang, Y.}, \bibinfo{author}{Li, Z.}, \bibinfo{author}{Zhang,
  H.}, \bibinfo{author}{Kong, Y.}, \bibinfo{author}{Wu, Y.},
  \bibinfo{author}{So, H. K.-H.}, \bibinfo{author}{Guo, Z.},
  \bibinfo{author}{Zhu, L.}, \& \bibinfo{author}{Wong, N.}
  (\bibinfo{year}{2025}).
\newblock \bibinfo{title}{{TreeReview: A dynamic tree of questions framework
  for deep and efficient LLM-based scientific peer review}}.
\newblock In {\it \bibinfo{booktitle}{EMNLP}\/}.
\newblock \DOIprefix\doi{10.18653/v1/2025.emnlp-main.790}.
\bibitem[{Couto et~al.(2024)Couto, Ho, Kumari, Rachmat, Khuong, Ullah \&
  Sun-Hosoya}]{couto-relevai-2024}
\bibinfo{author}{Couto, P.~H.}, \bibinfo{author}{Ho, Q.~P.},
  \bibinfo{author}{Kumari, N.}, \bibinfo{author}{Rachmat, B.~K.},
  \bibinfo{author}{Khuong, T. G.~H.}, \bibinfo{author}{Ullah, I.}, \&
  \bibinfo{author}{Sun-Hosoya, L.} (\bibinfo{year}{2024}).
\newblock \bibinfo{title}{{RelevAI-Reviewer: A benchmark on AI reviewers for
  survey paper relevance}}.
\newblock In {\it \bibinfo{booktitle}{CAp 2024 - Conférence sur
  l'Apprentissage Automatique}\/}.
\bibitem[{Demetrio et~al.(2025)Demetrio, Apruzzese, Grosse, Laskov, Lupu,
  Rimmer \& Widmer}]{demetrio-gen-2025}
\bibinfo{author}{Demetrio, L.}, \bibinfo{author}{Apruzzese, G.},
  \bibinfo{author}{Grosse, K.}, \bibinfo{author}{Laskov, P.},
  \bibinfo{author}{Lupu, E.}, \bibinfo{author}{Rimmer, V.}, \&
  \bibinfo{author}{Widmer, P.} (\bibinfo{year}{2025}).
\newblock \bibinfo{title}{{Gen-Review: A large-scale dataset of AI-generated
  (and human-written) peer reviews}}.
\newblock {\it \bibinfo{journal}{arXiv preprint arXiv:2510.21192}\/}, .
\bibitem[{Du et~al.(2024)Du, Wang, Zhao, Deng, Liu, Lou, Zou, Narayanan~Venkit,
  Zhang, Srinath, Zhang, Gupta, Li, Li, Wang, Liu, Liu, Gao, Xia, Xing,
  Jiayang, Wang, Su, Shah, Guo, Gu, Li, Wei, Wang, Cheng, Ranathunga, Fang, Fu,
  Liu, Huang, Blanco, Cao, Zhang, Yu \& Yin}]{du-llms-2024}
\bibinfo{author}{Du, J.}, \bibinfo{author}{Wang, Y.}, \bibinfo{author}{Zhao,
  W.}, \bibinfo{author}{Deng, Z.}, \bibinfo{author}{Liu, S.},
  \bibinfo{author}{Lou, R.}, \bibinfo{author}{Zou, H.~P.},
  \bibinfo{author}{Narayanan~Venkit, P.}, \bibinfo{author}{Zhang, N.},
  \bibinfo{author}{Srinath, M.}, \bibinfo{author}{Zhang, H.~R.},
  \bibinfo{author}{Gupta, V.}, \bibinfo{author}{Li, Y.}, \bibinfo{author}{Li,
  T.}, \bibinfo{author}{Wang, F.}, \bibinfo{author}{Liu, Q.},
  \bibinfo{author}{Liu, T.}, \bibinfo{author}{Gao, P.}, \bibinfo{author}{Xia,
  C.}, \bibinfo{author}{Xing, C.}, \bibinfo{author}{Jiayang, C.},
  \bibinfo{author}{Wang, Z.}, \bibinfo{author}{Su, Y.}, \bibinfo{author}{Shah,
  R.~S.}, \bibinfo{author}{Guo, R.}, \bibinfo{author}{Gu, J.},
  \bibinfo{author}{Li, H.}, \bibinfo{author}{Wei, K.}, \bibinfo{author}{Wang,
  Z.}, \bibinfo{author}{Cheng, L.}, \bibinfo{author}{Ranathunga, S.},
  \bibinfo{author}{Fang, M.}, \bibinfo{author}{Fu, J.}, \bibinfo{author}{Liu,
  F.}, \bibinfo{author}{Huang, R.}, \bibinfo{author}{Blanco, E.},
  \bibinfo{author}{Cao, Y.}, \bibinfo{author}{Zhang, R.}, \bibinfo{author}{Yu,
  P.~S.}, \& \bibinfo{author}{Yin, W.} (\bibinfo{year}{2024}).
\newblock \bibinfo{title}{{LLMs assist NLP researchers: Critique paper
  (meta-)reviewing}}.
\newblock In {\it \bibinfo{booktitle}{EMNLP}\/}.
\newblock \DOIprefix\doi{10.18653/v1/2024.emnlp-main.292}.
\bibitem[{Ebrahimi et~al.(2025)Ebrahimi, Sadeghian, Ghorbanpour, Arabzadeh,
  Salamat, Li, Le, Bashari \& Bagheri}]{ebrahimi-rottenreviews-2025}
\bibinfo{author}{Ebrahimi, S.}, \bibinfo{author}{Sadeghian, S.},
  \bibinfo{author}{Ghorbanpour, A.}, \bibinfo{author}{Arabzadeh, N.},
  \bibinfo{author}{Salamat, S.}, \bibinfo{author}{Li, M.}, \bibinfo{author}{Le,
  H.~S.}, \bibinfo{author}{Bashari, M.}, \& \bibinfo{author}{Bagheri, E.}
  (\bibinfo{year}{2025}).
\newblock \bibinfo{title}{{RottenReviews: Benchmarking review quality with
  human and LLM-based judgments}}.
\newblock In {\it \bibinfo{booktitle}{CIKM}\/}.
\newblock \DOIprefix\doi{10.1145/3746252.3761506}.
\bibitem[{Eger et~al.(2025)Eger, Cao, D'Souza, Geiger, Greisinger, Gross, Hou,
  Krenn, Lauscher \& Li}]{eger-transforming-2025}
\bibinfo{author}{Eger, S.}, \bibinfo{author}{Cao, Y.},
  \bibinfo{author}{D'Souza, J.}, \bibinfo{author}{Geiger, A.},
  \bibinfo{author}{Greisinger, C.}, \bibinfo{author}{Gross, S.},
  \bibinfo{author}{Hou, Y.}, \bibinfo{author}{Krenn, B.},
  \bibinfo{author}{Lauscher, A.}, \& \bibinfo{author}{Li, Y.}
  (\bibinfo{year}{2025}).
\newblock \bibinfo{title}{{Transforming science with large language models: A
  survey on AI-assisted scientific discovery, experimentation, content
  generation, and evaluation}}.
\newblock {\it \bibinfo{journal}{arXiv preprint arXiv:2502.05151}\/}, .
\bibitem[{Elsevier(2024)}]{elsevier-use-2024}
\bibinfo{author}{Elsevier} (\bibinfo{year}{2024}).
\newblock \bibinfo{title}{{The use of generative AI and AI-assisted
  technologies in the journal peer review process}}.
\newblock \URLprefix
  \url{https://www.elsevier.com/about/policies-and-standards/generative-ai-policies-for-journals#2-for-reviewers}.
\bibitem[{Faizullah et~al.(2024)Faizullah, Urlana \&
  Mishra}]{faizullah-limgen-2024}
\bibinfo{author}{Faizullah, A. R. B.~M.}, \bibinfo{author}{Urlana, A.}, \&
  \bibinfo{author}{Mishra, R.} (\bibinfo{year}{2024}).
\newblock \bibinfo{title}{{LimGen: Probing the LLMs for generating suggestive
  limitations of research papers}}.
\newblock In {\it \bibinfo{booktitle}{ECML PKDD}\/}.
\newblock \DOIprefix\doi{10.1007/978-3-031-70344-7_7}.
\bibitem[{Gao et~al.(2025)Gao, Ruan, Zhang, Gao, Liu \& Fu}]{gao-mmreview-2025}
\bibinfo{author}{Gao, X.}, \bibinfo{author}{Ruan, J.}, \bibinfo{author}{Zhang,
  Z.}, \bibinfo{author}{Gao, J.}, \bibinfo{author}{Liu, T.}, \&
  \bibinfo{author}{Fu, Y.} (\bibinfo{year}{2025}).
\newblock \bibinfo{title}{{MMReview: A multidisciplinary and multimodal
  benchmark for LLM-based peer review automation}}.
\newblock {\it \bibinfo{journal}{arXiv preprint arXiv:2508.14146}\/}, .
\bibitem[{Gao et~al.(2024)Gao, Brantley \& Joachims}]{gao-reviewer2-2024}
\bibinfo{author}{Gao, Z.}, \bibinfo{author}{Brantley, K.}, \&
  \bibinfo{author}{Joachims, T.} (\bibinfo{year}{2024}).
\newblock \bibinfo{title}{{Reviewer2: Optimizing review generation through
  prompt generation}}.
\newblock {\it \bibinfo{journal}{arXiv preprint arXiv:2402.10886}\/}, .
\bibitem[{Garg et~al.(2025)Garg, Prasad, Singhal, Kirtani, Mandal \&
  Kumar}]{grag-revieweval-2025}
\bibinfo{author}{Garg, M.~K.}, \bibinfo{author}{Prasad, T.},
  \bibinfo{author}{Singhal, T.}, \bibinfo{author}{Kirtani, C.},
  \bibinfo{author}{Mandal, M.}, \& \bibinfo{author}{Kumar, D.}
  (\bibinfo{year}{2025}).
\newblock \bibinfo{title}{{ReviewEval: An evaluation framework for AI-generated
  reviews}}.
\newblock In {\it \bibinfo{booktitle}{Findings of EMNLP}\/}.
\newblock \DOIprefix\doi{10.18653/v1/2025.findings-emnlp.1120}.
\bibitem[{IEEE(2024)}]{ieee-become-2024}
\bibinfo{author}{IEEE} (\bibinfo{year}{2024}).
\newblock \bibinfo{title}{{Become an IEEE reviewer}}.
\newblock \URLprefix
  \url{https://journals.ieeeauthorcenter.ieee.org/submit-your-article-for-peer-review/become-an-ieee-reviewer/}.
\bibitem[{Jewitt et~al.(2025)Jewitt, Bezemer \&
  O'Halloran}]{jewitt-introducing-2025}
\bibinfo{author}{Jewitt, C.}, \bibinfo{author}{Bezemer, J.}, \&
  \bibinfo{author}{O'Halloran, K.} (\bibinfo{year}{2025}).
\newblock {\it \bibinfo{title}{{Introducing multimodality}}\/}.
\newblock (\bibinfo{edition}{2nd} ed.).
\newblock \bibinfo{publisher}{Routledge}.
\bibitem[{Ji et~al.(2025)Ji, Deng, Xue, Jin, Ding, Gan, Fu, Wang \&
  Zhou}]{ji-aceparse-2025}
\bibinfo{author}{Ji, H.}, \bibinfo{author}{Deng, C.}, \bibinfo{author}{Xue,
  B.}, \bibinfo{author}{Jin, Z.}, \bibinfo{author}{Ding, J.},
  \bibinfo{author}{Gan, X.}, \bibinfo{author}{Fu, L.}, \bibinfo{author}{Wang,
  X.}, \& \bibinfo{author}{Zhou, C.} (\bibinfo{year}{2025}).
\newblock \bibinfo{title}{{AceParse: A comprehensive dataset with diverse
  structured texts for academic literature parsing}}.
\newblock In {\it \bibinfo{booktitle}{ICASSP}\/}.
\newblock \DOIprefix\doi{10.1109/ICASSP49660.2025.10889977}.
\bibitem[{Jung et~al.(2025)Jung, Pyeon, Heo \& Ahn}]{jung-what-2025}
\bibinfo{author}{Jung, S.}, \bibinfo{author}{Pyeon, G.}, \bibinfo{author}{Heo,
  I.}, \& \bibinfo{author}{Ahn, H.} (\bibinfo{year}{2025}).
\newblock \bibinfo{title}{{What drives paper acceptance? A process-centric
  analysis of modern peer review}}.
\newblock {\it \bibinfo{journal}{arXiv preprint arXiv:2509.25701}\/}, .
\bibitem[{Karishma et~al.(2023)Karishma, Rohatgi, Puranik, Wu \&
  Giles}]{karishma-acl-2023}
\bibinfo{author}{Karishma, Z.}, \bibinfo{author}{Rohatgi, S.},
  \bibinfo{author}{Puranik, K.~S.}, \bibinfo{author}{Wu, J.}, \&
  \bibinfo{author}{Giles, C.~L.} (\bibinfo{year}{2023}).
\newblock \bibinfo{title}{{ACL-FIG: A dataset for scientific figure
  classification}}.
\newblock In {\it \bibinfo{booktitle}{AAAI-SDU}\/}.
\bibitem[{Li et~al.(2024{\natexlab{a}})Li, Wang, Xu, Wang, Feng, Kong \&
  Liu}]{li-multimodal-2024}
\bibinfo{author}{Li, L.}, \bibinfo{author}{Wang, Y.}, \bibinfo{author}{Xu, R.},
  \bibinfo{author}{Wang, P.}, \bibinfo{author}{Feng, X.},
  \bibinfo{author}{Kong, L.}, \& \bibinfo{author}{Liu, Q.}
  (\bibinfo{year}{2024}{\natexlab{a}}).
\newblock \bibinfo{title}{{Multimodal ArXiv: A dataset for improving scientific
  comprehension of large vision-language models}}.
\newblock In {\it \bibinfo{booktitle}{ACL}\/}.
\newblock \DOIprefix\doi{10.18653/v1/2024.acl-long.775}.
\bibitem[{Li et~al.(2025)Li, Zhang, Gehringer, Xiao, Ding \&
  Chen}]{li-unveiling-2025}
\bibinfo{author}{Li, R.}, \bibinfo{author}{Zhang, H.},
  \bibinfo{author}{Gehringer, E.}, \bibinfo{author}{Xiao, T.},
  \bibinfo{author}{Ding, J.}, \& \bibinfo{author}{Chen, H.}
  (\bibinfo{year}{2025}).
\newblock \bibinfo{title}{{Unveiling the merits and defects of LLMs in
  automatic review generation for scientific papers}}.
\newblock In {\it \bibinfo{booktitle}{ICDM}\/}.
\newblock \DOIprefix\doi{10.1109/ICDM65498.2025.00146}.
\bibitem[{Li et~al.(2024{\natexlab{b}})Li, Huang, Zhuang, Shi, Cai, Xu, Wang,
  Zhang, Ke \& Cai}]{li-scilitllm-2024}
\bibinfo{author}{Li, S.}, \bibinfo{author}{Huang, J.}, \bibinfo{author}{Zhuang,
  J.}, \bibinfo{author}{Shi, Y.}, \bibinfo{author}{Cai, X.},
  \bibinfo{author}{Xu, M.}, \bibinfo{author}{Wang, X.}, \bibinfo{author}{Zhang,
  L.}, \bibinfo{author}{Ke, G.}, \& \bibinfo{author}{Cai, H.}
  (\bibinfo{year}{2024}{\natexlab{b}}).
\newblock \bibinfo{title}{{SciLitLLM: How to adapt LLMs for scientific
  literature understanding}}.
\newblock In {\it \bibinfo{booktitle}{Neurips Workshop FM4Science}\/}.
\bibitem[{Li et~al.(2024{\natexlab{c}})Li, Yang, Choi, Zhu, Hsieh, Kim, Lim,
  Ji, Lee \& Yan}]{li-mmsci-graduate-2024}
\bibinfo{author}{Li, Z.}, \bibinfo{author}{Yang, X.}, \bibinfo{author}{Choi,
  K.}, \bibinfo{author}{Zhu, W.}, \bibinfo{author}{Hsieh, R.},
  \bibinfo{author}{Kim, H.}, \bibinfo{author}{Lim, J.~H.}, \bibinfo{author}{Ji,
  S.}, \bibinfo{author}{Lee, B.}, \& \bibinfo{author}{Yan, X.}
  (\bibinfo{year}{2024}{\natexlab{c}}).
\newblock \bibinfo{title}{{MMSci: A dataset for graduate-level multi-discipline
  multimodal scientific understanding}}.
\newblock {\it \bibinfo{journal}{arXiv preprint arXiv:2407.04903}\/}, .
\bibitem[{Lin et~al.(2023{\natexlab{a}})Lin, Song, Zhou, Chen \&
  Shi}]{lin-automated-2023}
\bibinfo{author}{Lin, J.}, \bibinfo{author}{Song, J.}, \bibinfo{author}{Zhou,
  Z.}, \bibinfo{author}{Chen, Y.}, \& \bibinfo{author}{Shi, X.}
  (\bibinfo{year}{2023}{\natexlab{a}}).
\newblock \bibinfo{title}{{Automated scholarly paper review: Concepts,
  technologies, and challenges}}.
\newblock {\it \bibinfo{journal}{Information Fusion}\/},  {\it
  \bibinfo{volume}{98}\/}, \bibinfo{pages}{101830}.
  \DOIprefix\doi{10.1016/j.inffus.2023.101830}.
\bibitem[{Lin et~al.(2023{\natexlab{b}})Lin, Song, Zhou, Chen \&
  Shi}]{lin-moprd-2023}
\bibinfo{author}{Lin, J.}, \bibinfo{author}{Song, J.}, \bibinfo{author}{Zhou,
  Z.}, \bibinfo{author}{Chen, Y.}, \& \bibinfo{author}{Shi, X.}
  (\bibinfo{year}{2023}{\natexlab{b}}).
\newblock \bibinfo{title}{{MOPRD: A multidisciplinary open peer review
  dataset}}.
\newblock {\it \bibinfo{journal}{Neural Computing and Applications}\/},  {\it
  \bibinfo{volume}{35}\/}, \bibinfo{pages}{24191–24206}.
  \DOIprefix\doi{10.1007/s00521-023-08891-5}.
\bibitem[{Lu et~al.(2025{\natexlab{a}})Lu, Xu, Li, Ding \&
  Meng}]{lu-agent-2025}
\bibinfo{author}{Lu, K.}, \bibinfo{author}{Xu, S.}, \bibinfo{author}{Li, J.},
  \bibinfo{author}{Ding, K.}, \& \bibinfo{author}{Meng, G.}
  (\bibinfo{year}{2025}{\natexlab{a}}).
\newblock \bibinfo{title}{{Agent Reviewers: Domain-specific multimodal agents
  with shared memory for paper review}}.
\newblock In {\it \bibinfo{booktitle}{ICML}\/}.
\bibitem[{Lu et~al.(2025{\natexlab{b}})Lu, Kuznetsov \&
  Gurevych}]{lu-identifying-2025}
\bibinfo{author}{Lu, S.}, \bibinfo{author}{Kuznetsov, I.}, \&
  \bibinfo{author}{Gurevych, I.} (\bibinfo{year}{2025}{\natexlab{b}}).
\newblock \bibinfo{title}{{Identifying aspects in peer reviews}}.
\newblock In {\it \bibinfo{booktitle}{Findings of EMNLP}\/}.
\newblock \DOIprefix\doi{10.18653/v1/2025.findings-emnlp.326}.
\bibitem[{Markhasin(2025{\natexlab{a}})}]{markhasin-ai-2025}
\bibinfo{author}{Markhasin, E.} (\bibinfo{year}{2025}{\natexlab{a}}).
\newblock \bibinfo{title}{{AI-driven scholarly peer review via persistent
  workflow prompting, meta-prompting, and meta-reasoning}}.
\newblock {\it \bibinfo{journal}{arXiv preprint arXiv:2505.03332}\/}, .
\bibitem[{Markhasin(2025{\natexlab{b}})}]{markhasin-llm-2025}
\bibinfo{author}{Markhasin, E.} (\bibinfo{year}{2025}{\natexlab{b}}).
\newblock \bibinfo{title}{{LLM context conditioning and PWP prompting for
  multimodal validation of chemical formulas}}.
\newblock {\it \bibinfo{journal}{arXiv preprint arXiv:2505.12257}\/}, .
\bibitem[{Moosavi et~al.(2021)Moosavi, Rücklé, Roth \&
  Gurevych}]{moosavi-scigen-2021}
\bibinfo{author}{Moosavi, N.~S.}, \bibinfo{author}{Rücklé, A.},
  \bibinfo{author}{Roth, D.}, \& \bibinfo{author}{Gurevych, I.}
  (\bibinfo{year}{2021}).
\newblock \bibinfo{title}{{SciGen: A dataset for reasoning-aware text
  generation from scientific tables}}.
\newblock In {\it \bibinfo{booktitle}{NeurIPS 2021 Datasets and Benchmarks
  Track}\/}.
\bibitem[{Ng et~al.(2025)Ng, Maduranayagam, Suthakar, Li, Lokker, Iorio, Haynes
  \& Moher}]{ng-attitudes-2025}
\bibinfo{author}{Ng, J.~Y.}, \bibinfo{author}{Maduranayagam, S.~G.},
  \bibinfo{author}{Suthakar, N.}, \bibinfo{author}{Li, A.},
  \bibinfo{author}{Lokker, C.}, \bibinfo{author}{Iorio, A.},
  \bibinfo{author}{Haynes, R.~B.}, \& \bibinfo{author}{Moher, D.}
  (\bibinfo{year}{2025}).
\newblock \bibinfo{title}{{Attitudes and perceptions of medical researchers
  towards the use of artificial intelligence chatbots in the scientific
  process: an international cross-sectional survey}}.
\newblock {\it \bibinfo{journal}{The Lancet Digital Health}\/},  {\it
  \bibinfo{volume}{7}\/}, \bibinfo{pages}{e94–e102}.
  \DOIprefix\doi{10.1016/S2589-7500(24)00202-4}.
\bibitem[{Ni(2023)}]{ni-chatreviewer-2023}
\bibinfo{author}{Ni, S.} (\bibinfo{year}{2023}).
\newblock \bibinfo{title}{{ChatReviewer}}.
\newblock \URLprefix \url{https://github.com/nishiwen1214/ChatReviewer}.
\bibitem[{Song et~al.(2025{\natexlab{a}})Song, Xu, Wang, Wang \&
  Li}]{song-multidisciplinary-2025}
\bibinfo{author}{Song, H.}, \bibinfo{author}{Xu, H.}, \bibinfo{author}{Wang,
  Z.}, \bibinfo{author}{Wang, Y.}, \& \bibinfo{author}{Li, J.}
  (\bibinfo{year}{2025}{\natexlab{a}}).
\newblock \bibinfo{title}{{A multidisciplinary multimodal aligned dataset for
  academic data processing}}.
\newblock {\it \bibinfo{journal}{Scientific Data}\/},  {\it
  \bibinfo{volume}{12}\/}, \bibinfo{pages}{172}.
  \DOIprefix\doi{10.1038/s41597-025-04415-z}.
\bibitem[{Song et~al.(2025{\natexlab{b}})Song, Li, Li, Zhao, Yu, Ma, Mao, Zhang
  \& Wang}]{song-how-2025}
\bibinfo{author}{Song, S.}, \bibinfo{author}{Li, X.}, \bibinfo{author}{Li, S.},
  \bibinfo{author}{Zhao, S.}, \bibinfo{author}{Yu, J.}, \bibinfo{author}{Ma,
  J.}, \bibinfo{author}{Mao, X.}, \bibinfo{author}{Zhang, W.}, \&
  \bibinfo{author}{Wang, M.} (\bibinfo{year}{2025}{\natexlab{b}}).
\newblock \bibinfo{title}{{How to bridge the gap between modalities: Survey on
  multimodal large language model}}.
\newblock {\it \bibinfo{journal}{IEEE Transactions on Knowledge and Data
  Engineering}\/},  {\it \bibinfo{volume}{37}\/}, \bibinfo{pages}{5311–5329}.
  \DOIprefix\doi{10.1109/TKDE.2025.3527978}.
\bibitem[{{Springer Nature}(2024)}]{springer-editorial-2024}
\bibinfo{author}{{Springer Nature}} (\bibinfo{year}{2024}).
\newblock \bibinfo{title}{{Editorial policies - Artificial intelligence}}.
\newblock \URLprefix
  \url{https://www.springernature.com/gp/policies/editorial-policies}.
\bibitem[{Staudinger et~al.(2024)Staudinger, Kusa, Piroi \&
  Hanbury}]{staudinger-analysis-2024}
\bibinfo{author}{Staudinger, M.}, \bibinfo{author}{Kusa, W.},
  \bibinfo{author}{Piroi, F.}, \& \bibinfo{author}{Hanbury, A.}
  (\bibinfo{year}{2024}).
\newblock \bibinfo{title}{{An analysis of tasks and datasets in peer
  reviewing}}.
\newblock In {\it \bibinfo{booktitle}{SDP}\/}.
\bibitem[{Sundar et~al.(2024)Sundar, Xu, Gay, Richardson \&
  Heck}]{sundar-cpapers-2024}
\bibinfo{author}{Sundar, A.}, \bibinfo{author}{Xu, J.}, \bibinfo{author}{Gay,
  W.}, \bibinfo{author}{Richardson, C.}, \& \bibinfo{author}{Heck, L.}
  (\bibinfo{year}{2024}).
\newblock \bibinfo{title}{{cPAPERS: A dataset of situated and multimodal
  interactive conversations in scientific papers}}.
\newblock In {\it \bibinfo{booktitle}{NeurIPS}\/}.
\newblock \DOIprefix\doi{10.52202/079017-2119}.
\bibitem[{Taechoyotin et~al.(2024)Taechoyotin, Wang, Zeng, Sides \&
  Acuna}]{taechoyotin-mamorx-2024}
\bibinfo{author}{Taechoyotin, P.}, \bibinfo{author}{Wang, G.},
  \bibinfo{author}{Zeng, T.}, \bibinfo{author}{Sides, B.}, \&
  \bibinfo{author}{Acuna, D.} (\bibinfo{year}{2024}).
\newblock \bibinfo{title}{{MAMORX: Multi-agent multi-modal scientific review
  generation with external knowledge}}.
\newblock In {\it \bibinfo{booktitle}{NeurIPS Workshop FM4Science}\/}.
\bibitem[{Tan \& Rigger(2024)}]{tan-inconsistencies-2024}
\bibinfo{author}{Tan, J.}, \& \bibinfo{author}{Rigger, M.}
  (\bibinfo{year}{2024}).
\newblock \bibinfo{title}{{Inconsistencies in TeX-produced documents}}.
\newblock In {\it \bibinfo{booktitle}{ISSTA}\/}.
\newblock \DOIprefix\doi{10.1145/3650212.3680370}.
\bibitem[{Thakkar et~al.(2026)Thakkar, Yuksekgonul, Silberg, Garg, Peng, Sha,
  Yu, Vondrick \& Zou}]{thakkar-large-2026}
\bibinfo{author}{Thakkar, N.}, \bibinfo{author}{Yuksekgonul, M.},
  \bibinfo{author}{Silberg, J.}, \bibinfo{author}{Garg, A.},
  \bibinfo{author}{Peng, N.}, \bibinfo{author}{Sha, F.}, \bibinfo{author}{Yu,
  R.}, \bibinfo{author}{Vondrick, C.}, \& \bibinfo{author}{Zou, J.}
  (\bibinfo{year}{2026}).
\newblock \bibinfo{title}{{A large-scale randomized study of large language
  model feedback in peer review}}.
\newblock {\it \bibinfo{journal}{Nature Machine Intelligence}\/},  {\it
  \bibinfo{volume}{8}\/}, \bibinfo{pages}{326–336}.
  \DOIprefix\doi{10.1038/s42256-026-01188-x}.
\bibitem[{Yang et~al.(2025)Yang, Shi, Liu, Shui, Wang, Jing, Xu, Zhu, Li,
  Zhang, Liu, Nie, Cai \& Yang}]{yang-chartmimic-2025}
\bibinfo{author}{Yang, C.}, \bibinfo{author}{Shi, C.}, \bibinfo{author}{Liu,
  Y.}, \bibinfo{author}{Shui, B.}, \bibinfo{author}{Wang, J.},
  \bibinfo{author}{Jing, M.}, \bibinfo{author}{Xu, L.}, \bibinfo{author}{Zhu,
  X.}, \bibinfo{author}{Li, S.}, \bibinfo{author}{Zhang, Y.},
  \bibinfo{author}{Liu, G.}, \bibinfo{author}{Nie, X.}, \bibinfo{author}{Cai,
  D.}, \& \bibinfo{author}{Yang, Y.} (\bibinfo{year}{2025}).
\newblock \bibinfo{title}{{ChartMimic: Evaluating LMM's cross-modal reasoning
  capability via chart-to-code generation}}.
\newblock In {\it \bibinfo{booktitle}{ICLR}\/}.
\bibitem[{Yang et~al.(2024)Yang, Dabre, Tanaka \& Okazaki}]{yang-scicap-2024}
\bibinfo{author}{Yang, Z.}, \bibinfo{author}{Dabre, R.},
  \bibinfo{author}{Tanaka, H.}, \& \bibinfo{author}{Okazaki, N.}
  (\bibinfo{year}{2024}).
\newblock \bibinfo{title}{{SciCap+: A knowledge augmented dataset to study the
  challenges of scientific figure captioning}}.
\newblock {\it \bibinfo{journal}{Journal of Natural Language Processing}\/},
  {\it \bibinfo{volume}{31}\/}, \bibinfo{pages}{1140–1165}.
  \DOIprefix\doi{10.5715/jnlp.31.1140}.
\bibitem[{Yin et~al.(2024)Yin, Fu, Zhao, Li, Sun, Xu \& Chen}]{yin-survey-2024}
\bibinfo{author}{Yin, S.}, \bibinfo{author}{Fu, C.}, \bibinfo{author}{Zhao,
  S.}, \bibinfo{author}{Li, K.}, \bibinfo{author}{Sun, X.},
  \bibinfo{author}{Xu, T.}, \& \bibinfo{author}{Chen, E.}
  (\bibinfo{year}{2024}).
\newblock \bibinfo{title}{{A survey on multimodal large language models}}.
\newblock {\it \bibinfo{journal}{National Science Review}\/},  {\it
  \bibinfo{volume}{11}\/}. \DOIprefix\doi{10.1093/nsr/nwae403}.
\bibitem[{Yu et~al.(2024)Yu, Ding, Tan, Luo, Weng, Gong, Zeng, Cui, Han, Sun,
  Wu, Lan \& Li}]{yu-automated-2024}
\bibinfo{author}{Yu, J.}, \bibinfo{author}{Ding, Z.}, \bibinfo{author}{Tan,
  J.}, \bibinfo{author}{Luo, K.}, \bibinfo{author}{Weng, Z.},
  \bibinfo{author}{Gong, C.}, \bibinfo{author}{Zeng, L.}, \bibinfo{author}{Cui,
  R.}, \bibinfo{author}{Han, C.}, \bibinfo{author}{Sun, Q.},
  \bibinfo{author}{Wu, Z.}, \bibinfo{author}{Lan, Y.}, \& \bibinfo{author}{Li,
  X.} (\bibinfo{year}{2024}).
\newblock \bibinfo{title}{{Automated peer reviewing in paper SEA:
  Standardization, evaluation, and analysis}}.
\newblock In {\it \bibinfo{booktitle}{Findings of EMNLP}\/}.
\newblock \DOIprefix\doi{10.18653/v1/2024.findings-emnlp.595}.
\bibitem[{Zhu et~al.(2025)Zhu, Weng, Yang \& Zhang}]{zhu-deepreview-2025}
\bibinfo{author}{Zhu, M.}, \bibinfo{author}{Weng, Y.}, \bibinfo{author}{Yang,
  L.}, \& \bibinfo{author}{Zhang, Y.} (\bibinfo{year}{2025}).
\newblock \bibinfo{title}{{DeepReview: Improving LLM-based paper review with
  human-like deep thinking process}}.
\newblock In {\it \bibinfo{booktitle}{ACL}\/}.
\newblock \DOIprefix\doi{10.18653/v1/2025.acl-long.1420}.
\bibitem[{Zhuang et~al.(2025)Zhuang, Chen, Xu, Jiang \&
  Lin}]{zhuang-large-2025}
\bibinfo{author}{Zhuang, Z.}, \bibinfo{author}{Chen, J.}, \bibinfo{author}{Xu,
  H.}, \bibinfo{author}{Jiang, Y.}, \& \bibinfo{author}{Lin, J.}
  (\bibinfo{year}{2025}).
\newblock \bibinfo{title}{{Large language models for automated scholarly paper
  review: A survey}}.
\newblock {\it \bibinfo{journal}{Information Fusion}\/},  {\it
  \bibinfo{volume}{124}\/}, \bibinfo{pages}{103332}.
  \DOIprefix\doi{10.1016/j.inffus.2025.103332}.

\end{thebibliography}

\end{CJK}
\end{document}